\documentclass[12pt]{article}

\usepackage{setspace,graphicx,epstopdf,amsmath,amsfonts,amssymb,amsthm,versionPO}
\usepackage{marginnote,datetime,enumitem,subfigure,rotating,fancyvrb}
\usepackage{hyperref,float}
\usepackage[longnamesfirst]{natbib}
\usepackage{color}
\usepackage{colortbl}
\usepackage{multirow}
\usepackage{url}
\usepackage[official]{eurosym}
\usdate

\excludeversion{notes}		
\includeversion{links}          

\iflinks{}{\hypersetup{draft=true}}

\ifnotes{%
\usepackage[margin=1in,paperwidth=10in,right=2.5in]{geometry}%
\usepackage[textwidth=1.4in,shadow,colorinlistoftodos]{todonotes}%
}{%
\usepackage[margin=1in]{geometry}%
\usepackage[disable]{todonotes}%
}

\makeatletter\let\chapter\@undefined\makeatother 

\usepackage{indentfirst} 
\usepackage{jf}          
\usepackage[labelfont=bf,labelsep=period]{caption}   
\def\EMA{{\rm EMA}}
\def\EV{{\rm FCL}}

\def\N{n}
\def\Ns{{n_s}}

\def\Port{\pi}
\def\PP{{r_\Port}}

\def\etal{~{\it et al.}}

\begin{document}

\setlist{noitemsep}  
\onehalfspacing      

\author{Sebastien Valeyre\thanks{\rm John Locke Investments, 38 Avenue Franklin Roosevelt, 77210 Fontainebleau-Avon, France; 
and Universit\'e de Paris XIII, Sorbonne Paris Cit\'e , CEPN, UMR-CNRS 7234, 93430 Villetaneuse, France},
Denis Grebenkov\thanks{\rm Laboratoire de Physique de la Mati\`{e}re Condens\'{e}e (UMR 7643), CNRS -- Ecole Polytechnique, 91128 Palaiseau, France},
Sofiane Aboura\thanks{\rm Universit\'e de Paris XIII, Sorbonne Paris Cit\'e, CEPN, UMR-CNRS 7234, 93430 Villetaneuse, France},
and Francois Bonnin\thanks{\rm John Locke Investments, 38 Avenue Franklin Roosevelt, 77210 Fontainebleau-Avon, France}}

\title{\Large \bf Should employers pay their employees better? \\ An asset pricing approach}

\date{\today}              


\maketitle
\thispagestyle{empty}

\bigskip

\centerline{\bf ABSTRACT}

\begin{doublespace}  
We uncover a new anomaly in asset pricing that is linked to the
remuneration: the more a company spends on salaries and benefits per
employee, the better its stock performs, on average.  Moreover, the
companies adopting similar remuneration policies share a common risk,
which is comparable to that of the value premium.  For this purpose,
we set up an original methodology that uses firm financial
characteristics to build factors that are less correlated than in the
standard asset pricing methodology.  We quantify the importance of
these factors from an asset pricing perspective by introducing the
factor correlation level as a directly accessible proxy of eigenvalues
of the correlation matrix.  A rational explanation of the remuneration
anomaly involves the positive correlation between pay and employee
performance.
\end{doublespace}

\medskip




\clearpage


\onehalfspace

\section{Introduction}

Should employers pay their employees better?  Although this question
might appear provoking because lowering production costs remains a
cornerstone of the contemporary economy, we present the first attempt
to report the real effects of employee remuneration on asset pricing.
Remuneration -- defined as the annual salaries and benefits expenses
(e.g., wages, bonuses, pension expenses, health insurance payment,
etc.) per employee -- is the basis of any employment contract.  For
instance, pay was shown to explain, on average, 65\% of the variance
in evaluations of overall job attractiveness \citep{Rynes83}.
Classical theory states that profit-maximizing firms choose the level
of labor pay by setting the marginal cost of labor (i.e., the wage
rate) equal to the marginal revenue product of labor (i.e., the
marginal benefit). Beyond this paradigm, we provide strong evidence
that firms that pay their employees better tend to over-perform on the
stock market.

Our objective is to examine whether remuneration is an anomaly that
can be priced in asset pricing models.  \citet{Schwert03} defines
anomalies as ``empirical results that seem to be inconsistent with
maintained theories of asset-pricing behavior (the CAPM).  They
indicate either market inefficiency (profit opportunities) or
inadequacies in the asset-pricing model.  After they are documented
and analyzed in the academic literature, anomalies often seem to
disappear, reverse, or attenuate.''  Anomalies are typically
identified either by regressing a cross-section of average returns
(e.g., the seminal \citet{Fama73} approach uses the capitalization and
book-to-market values), or by using a panel regression of the
cross-section of returns with different factor returns through the
F-Statistic \citep{Gibbons89}, or by using a portfolio-based approach
that segregates individual stocks with similar capitalization and
book-to-market values into different style portfolios \citep{Fama93}.
In the latter case (which we refer to as the ``FF approach''), the
factors formed on small minus big market capitalization portfolios
(SMB) and high minus low book-to-market portfolios (HML) explain an
important part of the identified anomalies \citep{Fama96}.  Over
recent decades, the growing number of discovered anomalies suggests
that the standard asset pricing models fail to explain much of the
cross-sectional variation in average stock returns.  Meanwhile, the
effect of remuneration on company performance has surprisingly never
been tested, despite the fact that employers pay particular attention
to labor costs in attempting to maximize profits.

This research contributes empirically to the asset pricing literature
by introducing an observable firm characteristic, namely the
remuneration, as a candidate anomaly.  More precisely, we focus on
remuneration as a priced factor.  Indeed, it remains unclear how far
remuneration can explain the cross-section of returns despite a
sizeable literature on labor economics that relates labor to asset
pricing.  This branch of literature has intensively investigated the
impact of labor decisions on the firm's value, notably through the
operating leverage, which affects the equity returns riskiness.
However, to our best knowledge, there are no asset pricing studies
that incorporate employee's wages as a pricing factor.  Besides, based
on the impressive list of anomalies analyzed by \citet{Harvey15a}, we
find only one paper that highlights income as a potential factor.
Indeed, \citet{Gomez15} analyze the relation between U.S. census
division-level labor income and the cross-section of returns using the
standard \citet{Fama93} approach.  More specifically, these authors
use per capita personal income (from the Bureau of Economic Analysis)
as a new candidate factor and conclude that the cross-section of stock
returns depends on the census district in which the headquarters of
the firm are located.  Unfortunately, as \citet{Harvey15a} has noted,
``most of the division level labor income have a non-significant
t-statistic. \textsl{We do not count their factors}''.  Moreover, we
use remuneration at the company level to generate results that are
more realistic from an asset pricing perspective, which contrasts with
\citet{Gomez15}, whose scope is limited to income per state and per
division.

This research contributes also theoretically to the asset pricing
literature by introducing a new methodology to build factors that is
conceptually close to principal component analysis (PCA) but goes
beyond its noise-induced limitations.  This methodology presents many
advantages compared with the conventional multi-factor approach
developed by \citet{Fama92,Fama93}.  We propose a new measure of
``explanatory power'' of factors where the relevance of the factor
does not depend on the number of considered factors, in contrast to
the R-squared argument of the FF setting.  Hence, we introduce the
Factor Correlation Level (FCL) as a metrics of common risks that
measures the ability of stocks within the factor to fluctuate in a
common way.  Importantly, it allows ordering the factors according to
their capacity of taking into account the variability of stocks, and
therefore to their importance from an asset pricing perspective.  In
this respect, our ranking by the FCL indicator resembles principal
component analysis.  At the same time, this indicator is also linked
to the R-squared value of the factor in the asset pricing model:
higher FCLs correspond to higher R-squared values in the asset pricing
model with one factor.
The empirical validation of the FCL methodology is founded on an
exhaustive testing protocol.  First, we use ten factors that summarize
most of the existing factors: dividend, capitalization, liquidity,
momentum, low-volatility, debt-to-book, sales-to-market,
book-to-market, cash and, of course, the remuneration factor; those
which are not present in this list remain correlated with some of
these factors; we check that performance associated with the
remuneration factor is not explained by other major factors such as
low-volatility, capitalization, book-to-market, or momentum.  Second,
we consider six ``supersectors'' that are used to split stocks into
comparable groups since remuneration varies strongly from one sector
to another.  Third, we employ a large data set of 3612 daily single
stock close prices from January 2001 to July 2015 for the 569 biggest
companies in Europe.  For comparison, we also treat the same number of
randomly selected companies in the U.S.A. whose capitalization exceeds
1 billion of dollars.  Although we do not access the remuneration data
for these companies, the analysis of other factors allows us to
validate the FCL methodology on the U.S. market (often considered as a
benchmark) and to compare our predictions to whose of the FF approach.
Fourth, we perform several robustness checks to examine if the results
change with the tested variations; for instance, we perform a separate
analysis with the 258 biggest companies from U.K. to check for
potential domestic biases; we also run the methodology on monthly data
to check the role of time scale; in the spirit of comparability, we
evaluate the factor performances with seven incremental transitions
from the standard FF approach to our methodology.  Finally, we compare
our results with the basic PCA and illustrate its limitations.  Our
main result indicates that a market neutral investment strategy based
on the remuneration anomaly would likely deliver positive annual
returns of 2.42\% above the market.

The remainder of the paper is organized as follows.  Section
\ref{sec:literature} offers a literature review that covers several
fields of research.  Section \ref{sec:methodology} describes the novel
methodology.  Section \ref{sec:data} presents the data, whereas
Section \ref{sec:results} presents the empirical results.  Section
\ref{sec:discussion} discusses the advantages and limitations of our
methodology and compares it with the FF approach.  Section
\ref{sec:conclusion} summarizes the main findings and concludes.

\section{Literature review}
\label{sec:literature}

\subsection{The asset pricing}

This article is mainly related to the asset pricing literature in
which previous studies have shown that the average returns of common
stocks are related to firm characteristics such as capitalization,
price-earnings ratio, cash flow, book-to-market, past sales growth and
past returns. For example, stocks with lower market capitalization
tend to have higher average returns \citep{Banz81}.  Another important
anomaly is the value premium: value stocks have higher returns than
growth stocks, which is likely because the market undervalues
distressed stocks \citep{Fama98}.  More precisely, small stocks and
value stocks have higher average returns than their betas can explain
\citep{Campbell04}.  Profitability and investment also add to the
description of average returns \citep{Fama15}.  The low volatility
anomaly was revealed for medium and big stocks in addition to growth
stocks \citep{Jordan13}.  Those stocks that are expected to have high
idiosyncratic risk earn high returns in the cross-section
\citep{Fu09}.  This result contradicts previous findings made by
\citet{Ang06}, who posit that stocks with high idiosyncratic
volatility have low average returns.  Macroeconomic risk has also been
connected with the cross-section of returns.  For instance, the growth
rate of industrial production is seen as a priced risk factor in
standard asset pricing tests \citep{Chen86,Liu08}.  There is a size
effect in bank stock returns that differs from the market
capitalization effects documented in non-financial stock returns
\citep{Gandhi15}.  The most popular anomaly is momentum: stocks with
low past returns tend to have low future returns while stocks with
high past returns tend to have high future returns
\citep{Jegadeesh93}.  Hence, the momentum strategy that buys past
winners and sells past losers should earn abnormal returns in upcoming
years.  Return momentum has also been observed when spreads in average
momentum returns decrease from smaller to bigger stocks
\citep{Fama12}.  However, momentum strategies seem to produce losses
specifically in January \citep{Jegadeesh93}, probably based on
taxation effects \citep{Grinblatt04}.  Similarly, changes in book
equity appear to be more informative about expected stock returns than
price returns \citep{Bali13}.  Notably, certain stock market anomalies
may appear and then disappear after publication in academic journals
\citep{Mclean15}.  In spite of the abundant literature, the work by
\citet{Gomez15} seems to be the sole article that considers income as
a candidate anomaly although it is still not an income per employee
but rather per state and per division.
Several models have been developed to provide economic interpretations
of numerous stylized anomalies and to improve the performance of the
CAPM.%
\footnote{
\citet{Campbell04} introduced a two-beta model to explain the
capitalization and book-to-market value anomalies in stock returns by
splitting the CAPM into a cash-flow beta with a higher price of risk
than a discount-rate beta.  \citet{Fama93} proposed a three-factor
model to capture the patterns in U.S. average returns associated with
capitalization and value-versus-growth.  Even after a theoretical
rationale for the three-factor model was provided by
\citet{Ferguson03}, many anomalies remain unexplained by the
three-factor model \citep{Fama15}.  Although a four-factor model has
been derived \citep{Carhart97}, it has also failed to absorb all the
momentum in U.S. average stock returns \citep{Avramov06}.  Recently, a
five-factor model was introduced to capture capitalization, value,
profitability, and investment patterns in average stock returns and is
reputed to perform better than the three-factor model
\citep{Fama15}. }
Simultaneously, the anomaly-based evidence against the CAPM has been
questioned because anomalies have primarily been confined to small
stocks \citep{Cederburg15}.%
\footnote{ 
In line with this criticism, doubt was cast on the set of anomalies to
consider in a multi-factorial setup, given that \citet{Harvey15a} have
summarized 316 potential factors by reviewing 313 papers published
since 1967.  In the same vein, 38 out of 80 potential firm-level
anomalies were shown to be insignificant in the broad cross-section of
average stock returns \citep{Hou15}.  In addition, mistakes can easily
be made in this field due to multiple testing or data mining methods.
As noted by \citet{Harvey15b}, many discovered factors are likely to
be false if their t-statistics do not exceed 3.  Finally, these papers
suggest that many claims in the anomalies literature are likely to be
exaggerated regarding the associated t-statistics.}

\subsection{Corporate finance}

This article is also related to the extensive literature on corporate
finance, which has also continued to investigate the relation between
remuneration and performance, although it has usually focused on
managerial pay as opposed to the broader category of employees that we
consider in the present study.  This branch of literature typically
examines the wage as a managerial incentive likely to reduce agency
costs by designing an optimal job contract.  In that sense, we may
consider that solving the incentive problem leads to shareholder value
creation affecting stock returns.  Indeed, managers face both
discipline and opportunities provided by the free market economy that
leads to the notion that there is no need for explicit contracts to
resolve incentive problems \citep{Fama80}.  Nevertheless, market
forces cannot act as a complete substitute for contracts
\citep{Holmstrom99} because career concerns must be considered to
design optimal contracts and to arrive at strong incentives
\citep{Gibbons92}.  The effects of incentives depend on how they are
designed \citep{Gneezy11}, given that managers have considerable power
to shape their own pay arrangements -- and perhaps to even hurting
shareholder interest \citep{Bebchuk02}.  Indeed, public company
disclosures do not provide a comprehensive measure of managerial
incentive to increase shareholder value \citep{OByrne10}.  Many
explanations were brought forward to justify top managers'
remuneration.  Firms with abundant investment opportunities pay their
executives better \citep{Gaver95}. The increase in the level of
stock-option compensation can be explained by the inability of boards
to evaluate its real costs \citep{Hall03,Jensen04}.  The
capitalization of large firms explains many patterns in top manager
pay across firms, over time, and between countries \citep{Gabaix08}.
Manager fixed effects, interpreted as unobserved managerial attributes
and understood as a proxy for latent managerial ability, are important
in explaining the level of executive remuneration \citep{Graham12}.
Overall, remuneration matters because it may affect a corporation's
level of risk as bonus-driven remuneration might encourage excessive
risk-taking.  However, pay and risk are correlated not because
mis-aligned pay drives risk-taking, but rather because principal agent
theory predicts that riskier but more profitable firms must pay more
remuneration than less risky firms to provide a risk-averse manager
the same incentives \citep{Cheng15}.

\subsection{Labor economics}

The labor economics literature treats this question through the
``efficiency wage theory'' by relating it to unemployment.
\citet{Yellen84} and \citet{Akerlof90} did a remarkable work with an
analysis that is built -- unlike most economic models -- mainly on
sociology and psychology with experimentation that delivers salient
stylized facts on human behavior in a working context.  Efficiency
wage theory maintains that rising wages is the best way to increase
output per employee because it links pecuniary incentives to employee
performance.  In particular, the use of performance pay packages by
employers has been shown to increase employee productivity
\citep{Lazear00} and job satisfaction \citep{Green08}.  There are
several interesting studies that relate labor market to asset pricing.
All these empirical results emphasize the significant impact of labor
decisions, in which wage plays a prominent role, onto firm's value.
\citet{Santos06} show that labor income to consumption ratio is a
strong predictor of long horizon returns.  \citet{Danthine02} explain
that operating leverage is more significant for the riskiness of
equity returns than financial leverage.  In other words, attention
should be paid to wages, particularly because the priority nature of
wages enhances the risk of dividends.  In this spirit, \citet{Kuehn13}
note that a high value of unemployment makes wages inelastic, which
gives rise to operating leverage.  The impact of inelastic wages is
even stronger in bad times as it amplifies the equity risk premium.
\citet{Gourio07} argues that because wages are smooth, revenues are
more cyclic than costs, making the profits more volatile.  In
particular, firms with high book-to-market or with low productivity,
i.e. value firms, have more pro-cyclic earnings.  \citet{Ochoa13}
finds a positive and statistically significant relation between the
reliance on skilled labor and expected returns.  In times of high
volatility, firms with a high share of skilled workers earn an annual
return of 2.7\% above those with a high share of unskilled workers
notably because their labor is more costly to adjust.  Labor decisions
made by workers can affect firm risk \citep{Donangelo14} while hiring
decisions can also be the determinants of firm risk
\citep{Carlson04,Belo14}.  Indeed, \citet{Donangelo14} discusses the
idea that mobile workers carry some of the firm's capital productivity
when they leave an industry.  He finds that portfolios that hold long
positions in stocks of high-mobility industries (general workers) and
short positions in stocks of low-mobility industries
(industry-specific workers) earn an annual return spread of over 5\%.
Like \citet{Monika07} who explain that labor should matter since
firms' market value embodies the value of hiring, \citet{Belo14} argue
that the market value of a firm reflects the value of its labor force
because the firm can extract rents as compensation for the costs
associated with adjusting its labor force.  They find that long
positions in stocks of low-hiring firms and short positions in
high-hiring firms earn an average annual excess stock return of 5.6\%.
\citet{Favilukis16} introduce infrequent renegotiation in standard
wages model showing that it leads to smooth average wages.  Due to
this wage rigidity, they find that wage growth forecasts long-horizon
excess equity returns.

\subsection{Social sciences}

This article is also broadly related to several streams of research in
various social sciences, including sociology, psychology and human
resources.  In these fields, wage acts like a motivator since it
typically reflects a social preference for rewards likely to affect
the employee's performance.  Sociological studies have developed a
theory of social exchange in which there are equivalent rewards on
both sides \citep{Blau55}, which is consistent with the preference for
reciprocity that is viewed as a social preference, as it depends on
the behavior of the reference person \citep{Fehr02}.  Reciprocity
induces agents to cooperate voluntarily with the principal when the
principal treats them correctly; the evidence for reciprocity is based
on a so-called gift exchange experiment.

Psychological studies highlight the exchange in working situations in
which the perceived value of labor equals the perceived value of
remuneration, based on the theory of equity \citep{Adams63}. When
there is no mismatch between effort and wages, employees may change
their perceived effort and even their perceived level of remuneration
by redefining the non-pecuniary component.

Human resources studies generally offer evidence that money is an
important motivator for most people \citep{Rynes04}, as pay can help
climbing on the Maslow's motivational hierarchy of needs, including
social esteem and self-actualization.  Nevertheless, tangible rewards
might also produce secondary negative effects on motivation
\citep{Baker92} by forestalling self-regulation \citep{Deci99}.

\section{Methodology}
\label{sec:methodology}

In this section, we introduce a new methodology to build factors that
combines advantages of the PCA and the \citet{Fama93} approach.  As
would be the case with the PCA, our factors are built to be
uncorrelated with the market index and with sectorial factors.  For
each factor, {\bf we introduce and estimate the Factor Correlation
Level (FCL) that allows us to order the factors based on their
importance and to select the most important ones in asset pricing
models.}

\subsection{Conventional diagonalization of the covariance and correlation matrices}

Identifying common risks of multiple assets is necessary to diversify
investments and can help to profit from style's arbitrage
opportunities.  Conventional approaches, such as PCA, attempt to
diagonalize the empirical covariance (or correlation) matrix of the
traded universe, i.e., to decorrelate assets by constructing
independent linear combinations (portfolios) of assets.  Each
eigenvector of the covariance matrix represents the coefficients of
one such combination while the corresponding eigenvalue gives its
variance.  If the covariance matrix does not contain negative elements
(i.e., if there are no negatively correlated assets), the eigenvector
corresponding to the largest eigenvalue has positive elements that can
be interpreted as relative weights of stocks in the market mode. The
classical long portfolio, following the market, can be constructed by
investing in proportion to these weights.  In turn, market neutral
portfolios should be orthogonal to the market mode and therefore have
both long and short positions (the latter corresponding to negative
weights).  The other eigenvectors capture different common risks of
the traded universe, and the most common include sectorial risks
(e.g., banking sector, commodities, energy, etc.).

In mathematical terms, if the covariance matrix $\Omega$ of stocks was
known precisely, it might be diagonalized to identify uncorrelated
linear combinations of stocks and their variances to assess the
related risks.  For a traded universe with $\N$ stocks, let
$r_1,\ldots,r_n$ denote the daily returns of these stocks at a given
time.  The covariance matrix has $\N$ eigenvalues $\lambda_1, \ldots,
\lambda_\N$ and $\N$ eigenvectors $V_1,\ldots,V_\N$ satisfying $\Omega
V_\alpha = \lambda_\alpha V_\alpha$ (for each $\alpha = 1,\ldots,
\N$).  Each eigenvector $V_\alpha$ determines one linear combination
of stocks, $(V_\alpha)_1 r_1 + \ldots + (V_\alpha)_\N r_\N$, which is
decorrelated from the others, while the eigenvalue $\lambda_\alpha$ is
its variance (under the condition that $V_\alpha$ is appropriately
normalized).

The above eigenbasis can be interpreted as follows.  For any linear
combination of stocks with weights $w_i$, $\PP = w_1 r_1 + \ldots +
w_\N r_\N = (w \cdot r)$ (written as a scalar product), the variance
of such a portfolio $\Port$ can be expressed as
\begin{equation}
\langle \PP^2 \rangle = \langle \left(\sum\limits_{i=1}^\N w_i r_i\right)^2 \rangle =
\sum\limits_{i,j=1}^\N w_{i} w_{j} \Omega_{i,j} 
= \sum\limits_{i,j=1}^\N w_{i} w_{j}
\sum\limits_{\alpha=1}^\N \lambda_\alpha (V_\alpha)_i (V_\alpha)_j = \sum\limits_{\alpha=1}^\N \lambda_\alpha (w \cdot V_\alpha)^2 ,
\end{equation}
where $\langle \ldots \rangle$ denotes the expectation, and the
returns $r_k$ were assumed to be centered.  In other words, the
variance is decomposed into a sum of variances $\lambda_\alpha$ of
independent linear combinations proportional to the projection of the
weights $w_i$ onto the corresponding eigenvector $V_\alpha$.  If the
weights $w_i$ are chosen in proportion to the elements of one
eigenvector, i.e., $w_{i} = c (V_\alpha)_i$ for some $\alpha$ and $c$,
then the orthogonality of $V_\alpha$ to other eigenvectors yields
\begin{equation}
\langle \PP^2 \rangle = \lambda_\alpha c^2(V_\alpha \cdot V_\alpha)^2 = \lambda_\alpha ~ (w\cdot w),
\end{equation}
where we used the $L^2$-normalization of the eigenvectors: $(V_\alpha
\cdot V_\alpha) = 1$.  As expected, the variance of such a linear
combination is fully determined by the corresponding eigenvalue
$\lambda_\alpha$.  Notably, the above relation can also be written as
\begin{equation}
\lambda_\alpha = \frac{\langle \PP^2\rangle}{\sum\nolimits_{i=1}^\N w^2_{i}}
\end{equation}
to estimate the variance of the linear combination whose weights are
constructed close to an eigenvector.

As different stocks exhibit quite distinct volatilities, it is
convenient to rescale the stock's return $r_i$ by its realized
volatility $\sigma_i$: $\tilde{r}_i = r_i/\sigma_i$.  This rescaling
is also known to reduce heterogeneity of volatilities among stocks and
heteroskedasticity \citep{Andersen00,Bouchaud01b,Valeyre13}.  In other
words, one can write
\begin{equation}
\langle \PP^2 \rangle = \langle \left(\sum\limits_{i=1}^\N w_i \sigma_i \tilde{r}_i\right)^2 \rangle =
\sum\limits_{i,j=1}^\N \tilde{w}_i \tilde{w}_j C_{i,j}  ,
\end{equation}
where $\tilde{w}_i = w_i \sigma_i$ and $C = \langle \tilde{r}_i
\tilde{r}_j\rangle$ is the covariance matrix of the renormalized
returns $\tilde{r}_i$ or, equivalently, the correlation matrix of
returns $r_i$: $\Omega_{i,j} = \sigma_i \sigma_j C_{i,j}$.  To
proceed, the eigenvalues and eigenvectors of $\Omega$ can be replaced
by the eigenvalues $\tilde{\lambda}_\alpha$ and eigenvectors
$\tilde{V}_\alpha$ of the correlation matrix $C$, $C \tilde{V}_\alpha
= \tilde{\lambda}_\alpha \tilde{V}_\alpha$, i.e.,
\begin{equation}
\langle \PP^2 \rangle = \sum\limits_{i,j=1}^\N \tilde{w}_i \tilde{w}_j
\sum\limits_{\alpha=1}^\N \tilde{\lambda}_\alpha (\tilde{V}_\alpha)_i (\tilde{V}_\alpha)_j = 
\sum\limits_{\alpha=1}^\N \tilde{\lambda}_\alpha (\tilde{w} \cdot \tilde{V}_\alpha)^2 . 
\end{equation}
If the volatility-normalized weights $\tilde{w}_i$ are chosen to be
proportional to the elements of an eigenvector, $\tilde{w}_i =
c(\tilde{V}_\alpha)_i$, one obtains $\langle \PP^2 \rangle =
\tilde{\lambda}_\alpha c^2 (\tilde{V}_\alpha \cdot \tilde{V}_\alpha) =
\tilde{\lambda}_\alpha c^2 = \tilde{\lambda}_\alpha
(\tilde{w},\tilde{w})$, from which
\begin{equation}
\label{eq:lambdat_beta}
\tilde{\lambda}_\alpha = \frac{\langle \PP^2\rangle}{\sum\nolimits_{i=1}^\N w^2_i \sigma_i^2} ,
\end{equation}
where the $L^2$-normalization of $\tilde{V}_\alpha$ was used:
$(\tilde{V}_\alpha \cdot \tilde{V}_\alpha) = 1$.  As previously
discussed, $\tilde{\lambda}_\alpha$ is the rescaled variance of the
linear combination of the volatility-normalized returns $\tilde{r}_i$
(given by the eigenvector $\tilde{V}_\alpha$), each of which is
decorrelated from other such combinations.  By construction, the
variance $\tilde{\lambda}_\alpha$ is normalized, which facilitates the
comparison of different factors and different markets.  We emphasize
that diagonalizations of the covariance and correlation matrices are
generally not equivalent; in particular, the eigenvalues
$\lambda_\alpha$, $\tilde{\lambda}_\alpha$ and the eigenvectors
$V_\alpha$, $\tilde{V}_\alpha$ are different (though in our case,
their interpretations should be close).  We choose the second option
(i.e., Eq. (\ref{eq:lambdat_beta})), which inherently reduces stock
heterogeneity and heteroskedasticity due to rescaling.

Unfortunately, a straightforward diagonalization of the empirical
covariance or correlation matrix estimated from stock price series is
known to be very sensitive to noise
\citep{Laloux99,Plerou99,Plerou02,Potters05,Wang11,Allez12}.  In
particular, only a few eigenvectors corresponding to the largest
eigenvalues can be estimated, as illustrated and further discussed in
Sec. \ref{sec:PCA}.  As a consequence, conventional diagonalization
does not appear suitable for building various representative factors.

\subsection{Our methodology: Indicator-based factors}
\label{sec:theory_factors}

We propose a different approach to building factors.  We begin from
the available economic and financial indicators regarding the traded
companies, such as their capitalization, sales-to-market, dividend
yields, etc.  We expect that companies with comparable indicators --
at least those with comparable indicators in the extreme quantiles of
the indicator distribution -- will exhibit correlations in their stock
performance.  This hypothesis allows us to construct and then test
indicator-based factors beyond sectors.  To minimize sectorial
correlations, we split the stocks into six supersectors of similar
sizes, as detailed in Appendix \ref{sec:supersectors}.  The following
construction is performed separately for each supersector and then the
data are aggregated (see below).

We consider ten indicator-based factors:
\begin{enumerate}
\item
The dividend factor, which is based on the dividend yield.

\item
The capitalization (or size) factor, which is based on capitalization.

\item
The liquidity factor, which is based on the ratio of the weekly
exponential moving average to the total number of shares (i.e.,
capitalization/close price).

\item
The momentum factor, which is based on the 3-year exponential moving
average of past daily returns.

\item
The low-volatility (or beta) factor, which is based on the sensitivity
to the stock index.

\item
The leverage factor, which is based on the debt-to-book value ratio.

\item
The sales-to-market factor, which is based on the ratio of sales to
market value at the end of the fiscal period.

\item
The book-to-market factor, which is based on the ratio of the book
value to the market value at the end of the fiscal period.

\item
The remuneration factor, which is based on salaries and benefits
expense per employee.

\item
The cash factor, which is based on the ratio between the free cash
flow and the latest market value.

\end{enumerate}

We believe that considering these 10 factors is sufficient and
including additional factors will not significantly change our
results.  In particular we might have included the investment and
profitability factors following \citet{Fama15}, but we expect that our
10 factors already capture the common risk from these two factors.
Indeed, sales and cash should be correlated with profitability,
whereas the dividend yield and leverage ratio should be correlated
with investment.

For each trading day, the stocks of the chosen supersector are sorted
according to the indicator (e.g., remuneration) available the day
before (we use the publication date and not the valuation date).  The
related indicator-based factor is formed by buying the first $q\Ns$
stocks in the sorted list and shorting the last $q\Ns$ stocks, where
$\Ns$ is the number of stocks in the considered supersector, and $0 <
q < \frac12$ is a chosen quantile level.  The other stocks (with
intermediate indicator values) are not included (weighted by $0$).  In
the simplest setting, one can choose equal weights:
\begin{equation}
w_{i} = \left\{ \begin{array}{c l}   +1, & \qquad \textrm{if $i$ belongs to the first $q\Ns$ stocks in the sorted list}, \\
-1, & \qquad \textrm{if $i$ belongs to the last $q\Ns$ stocks in the sorted list}, \\
0, & \qquad \textrm{otherwise.} \\  \end{array}  \right.
\end{equation}
In attempting to reduce the specific risk, the common practice
suggests to invest inversely proportional to the stock's volatility
$\sigma_i$, i.e., to set $w_i = \pm 1/\sigma_i$ or $0$.  Moreover, the
inverse stock volatility should also be bounded to reduce the impact
of extreme specific risk.  Each trading day, we recompute the weight
$w_{i}$ as follows
\begin{equation}
\label{eq:weights}
w_{i} = \left\{ \begin{array}{c l}  + \mu_+ \min\{1, \sigma_{\rm mean}/\sigma_i  \}, & 
\qquad \textrm{if $i$ belongs to the first $q\Ns$ stocks in the sorted list}, \\
- \mu_- \min\{1, \sigma_{\rm mean}/\sigma_i \}, & 
\qquad  \textrm{if $i$ belongs to the last $q\Ns$ stocks in the sorted list},\\
0, & \qquad \textrm{otherwise,} \\  \end{array}  \right.
\end{equation}
where $\sigma_{\rm mean} = \frac{1}{\Ns}(\sigma_1 + \ldots +
\sigma_\Ns)$ is the mean estimated volatility over the supersector. In
this manner, the weights of low-volatility stocks are reduced to avoid
strongly unbalanced portfolios concentrated in such stocks.  The two
common multipliers, $\mu_\pm$, are used to ensure the beta market
neutral condition:
\begin{equation}
\label{eq:beta_neutral}
\sum\limits_{i=1}^\Ns \beta_{i} w_{i} = 0 ,
\end{equation}
where $\beta_{i}$ is the sensitivity of stock $i$ to the market
(obtained by a linear regression of the normalized stock and index
returns based on the reactive volatility model \citep{Valeyre13}; note
that the use of standard daily returns leads to similar results, see
Appendix \ref{sec:FF}).  If the aggregated sensitivity of the long
part of the portfolio to the market is higher than that of the short
part of the portfolio, its weight is reduced by the common multiplier
$\mu_+ < \frac{1}{2qn_s}$, which is obtained from
Eq. (\ref{eq:beta_neutral}) by setting $\mu_- = \frac{1}{2qn_s}$
(which implies that the sum of absolute weights $|w_i|$ does not
exceed $1$). In the opposite situation (when the short part of the
portfolio has a higher aggregated beta), one sets $\mu_+ =
\frac{1}{2qn_s}$ and determines the reducing multiplier $\mu_- <
\frac{1}{2qn_s}$ from Eq. (\ref{eq:beta_neutral}).  This method of
ensuring the market neutral condition is better than leaving the
residual beta (as in the FF approach) or withdrawing it by subtracting
an appropriate constant from all weights.  Indeed, under our approach,
the factor is maintained to be invested only in stocks that are
sensitive to this factor.  In turn, subtracting a constant would
affect all stocks, even those that were ``excluded'' and whose weights
were set to 0 in Eq. (\ref{eq:weights}).  We also emphasize the
difference with the conventional FF approach: {\bf our factors are
built to be market-neutral under Eq. (\ref{eq:beta_neutral}), whereas
the FF portfolio is built to be delta-neutral} (i.e., to have zero net
investment):
\begin{equation}
\sum\limits_{i=1}^\Ns w_{i} = 0 .
\end{equation}

The resulting factor is obtained by aggregating the weights
constructed for each supersector.  This construction is repeated for
each of the ten factors listed above.  We emphasize that the factors
are constructed on a daily basis, i.e., the weights are re-evaluated
daily based on updated indicators.  However, most indicators do not
change frequently so that the transaction costs related to changing
the factors are not significant.

The above procedure can be extended to construct factors from other
quantiles, in addition to the first and the last.  In this manner, we
will consider three portfolios for each factor:
\begin{itemize}
\item
Q1: long positions for stocks whose indicator belongs to the first
15\% quantile and short positions for stocks in the last 15\%
quantile, as discussed above (for $q = 0.15$).

\item
Q2: long positions for stocks in the second 15\% quantile and short
positions for stocks in the next-to-last 15\% quantile (i.e., positive
weights are assigned to stocks ranging between $0.15\Ns$ and $0.30\Ns$
in the list, and negative weights are assigned to stocks ranging
between $0.70\Ns$ and $0.85\Ns$).

\item
Q3: long positions for stocks in the third 15\% quantile ($0.30 \Ns -
0.45\Ns$) and short positions for stocks in the third-to-last 15\%
quantile ($0.55\Ns - 0.70\Ns$).
\end{itemize}

To evaluate common risk with each factor, we introduce the {\it factor
correlation level} (FCL) as the square root of the ratio between the
empirical variance of the indicator-based factor and the total
empirical variance of the constituent stocks:
\begin{equation}
\label{eq:EV_q}
\EV(t) = \left(\frac{\EMA\left\{ \PP^2(t) \right\}}{\EMA\left\{ \sum_{i=1}^{\N}{w_{i}^2(t) \sigma_i^2(t)} \right\}} \right)^{1/2},
\end{equation}
where $\PP(t)$ is the daily return of the factor,
\begin{equation}
\PP(t) = \sum\limits_{i=1}^\N w_{i}(t) r_i(t) ,
\end{equation}
where $w_{i}(t)$ is the weight of the stock $i$ in the factor, and
$\sigma_i(t)$ is the volatility of the stock $i$ estimated using the
reactive volatility model \citep{Valeyre13}.  The exponential moving
average (EMA) is used with a long averaging period of 200 days to
reduce noise by smoothing measurements.  We emphasize that the above
sum aggregates stocks from all supersectors.  We also considered the
standard volatility estimator based on a 40-days exponential moving
average and obtained similar results (see Appendix \ref{sec:FF}).  The
square root in Eq. (\ref{eq:EV_q}) is taken to operate with
volatilities instead of variances.  The estimator (\ref{eq:EV_q}) is
built analogously to Eq. (\ref{eq:lambdat_beta}) for the eigenvalues
$\tilde{\lambda}_\alpha$ of the correlation matrix.  This analogy
relies on the assumption that the indicator-based weights $w_{i}$ are
close to an eigenvector of the correlation matrix.  Since the true
correlation matrix is unavailable, it is impossible to directly
validate this strong assumption.  We will therefore resort to indirect
validations based on empirical correlations of the constructed factors
and on the profitability of trading strategies derived from such
factors.  Note also that the weights $w_{i}$ depend on the choice of
the quantile $q$, such that we expect to have slightly different
results for different quantiles (see Fig. \ref{fig:Remu} below).
Simultaneously, the analogy to eigenvalues of the correlation matrix
allows various factors to be classified according to their
``importance'': larger values of $\EV$ mean stronger volatility of the
factor and therefore higher common risks.  For example, when the
correlation of small capitalization firms increases while the
volatility of individuals stocks remains stable, the FCL of the
capitalization factor will increase, and the volatility of the factor
will increase.  In general, the risk of a factor is proportional to
the average individual volatility multiplied by the FCL.  For this
reason, {\bf FCL can be interpreted as an average correlation measure
between stocks within the factor that is also directly linked to the
common risk level underpinning the factor.}  It must also be
emphasized that the $\EV$ estimator is dynamic, i.e., it can capture
changes in the correlation structure of the market over time.

\section{Data}
\label{sec:data}

In this study, we use only liquid stocks (most with capitalization
greater than 800 million euros), thus excluding microcap firms that
are typically the main focus of the labor studies we have cited.
Thanks to the European accounting regulations, the remuneration must
be provided by European companies on a regular basis and can thus be
accessed through commercial databases such as FACTSET.  Lacking such
information for the U.S. market, we mainly focus on the European
companies.  To reveal possible nation-specific features, the analysis
is performed for two trading universes: (i) the 569 biggest companies
in Europe (London Stock Exchange, Euronext, Eurex, Sixt), and (ii) the
258 biggest companies on the London Stock Exchange only.  Although the
twice-as-large European universe is expected to increase the
statistical significance of the results, the consideration of the
U.K.-bounded universe allows us to eliminate country biases and
additional fluctuations (e.g., due to currency exchange rate
variations).  We will show that the major conclusions are similar for
both universes.  In addition, we will validate our indicator-based
methodology on the U.S. universe that includes the 569 randomly
selected companies whose capitalization is above 1 billions of dollar.
Note that the universe of the 1229 biggest firms in the U.S. studied
by \citet{Fama08} is comparable to our European universe in terms of
capitalization and liquidity.

All the companies that we include in the European and U.K. universes
belong to the small (below 1 billion euros), mid (between 1 and 5
billion euros), large (between 5 and 20 billion euros), or big (above
20 billion euros) capitalization categories.  The data set consists of
3612 daily single stock close prices from January 2001 to July 2015.
{\bf Note that most Fama and French data begin from 1963, which leads
to greater t-statistics.}  We rely on daily prices (instead of the
monthly prices that are commonly used in the literature) to have more
precision in the temporal granularity of our FCL estimation.  In
addition, several economic and financial indicators are extracted from
the FACTSET database: book-to-market, capitalization, sales-to-market,
dividend yield, debt-to-book, free cash flow, salaries and benefit
expenses, and the number of employees on an annual basis (see Table
\ref{tab:stats}).  For the European universe, we partly offset
geographical biases in each indicator by renormalizing it to its
median in the country.  For instance, remuneration is divided by its
median by country, whereas the median by country is subtracted from
the moving average of returns in the case of momentum.

\begin{table}
\begin{center}
\begin{tabular}{| c | c | c | c |} \hline
       & Capitalization & Number of employees & Remuneration \\  \hline
Europe & $(13 \pm 25)$ B\euro{} & $(41\pm 78)$ thousand & $(0.13 \pm 0.99)$ M\euro{} \\  \hline
U.K.   & $(11 \pm 21)$ B\pounds & $(38\pm 87)$ thousand & $(0.08 \pm 0.08)$ M\pounds \\  \hline
\end{tabular}
\end{center}
\caption{
~Basic statistics (mean and standard deviation) regarding
capitalization (in billions of euro/pounds), number of employees (in
thousands), and remuneration (in millions of euro/pounds) from the
FACTSET database.  Since minimal capitalization is approximately 800
million euros, the distribution is truncated at small
capitalizations. } 
\label{tab:stats}
\end{table}

\section{Empirical results}
\label{sec:results}

In this section, we present the main results of our methodology
applied to the European, the U.K. and the U.S. universes.  We mainly
focus on the remuneration indicator, which has largely been ignored so
far.  We will show that remuneration yields a non-negligible common
risk and represents a small anomaly.  The possibility of revealing the
role of the remuneration factor relies on the proposed FCL
methodology.

\subsection{Correlation between remuneration and capitalization}

\begin{figure}
\begin{center}
\includegraphics[width=120mm]{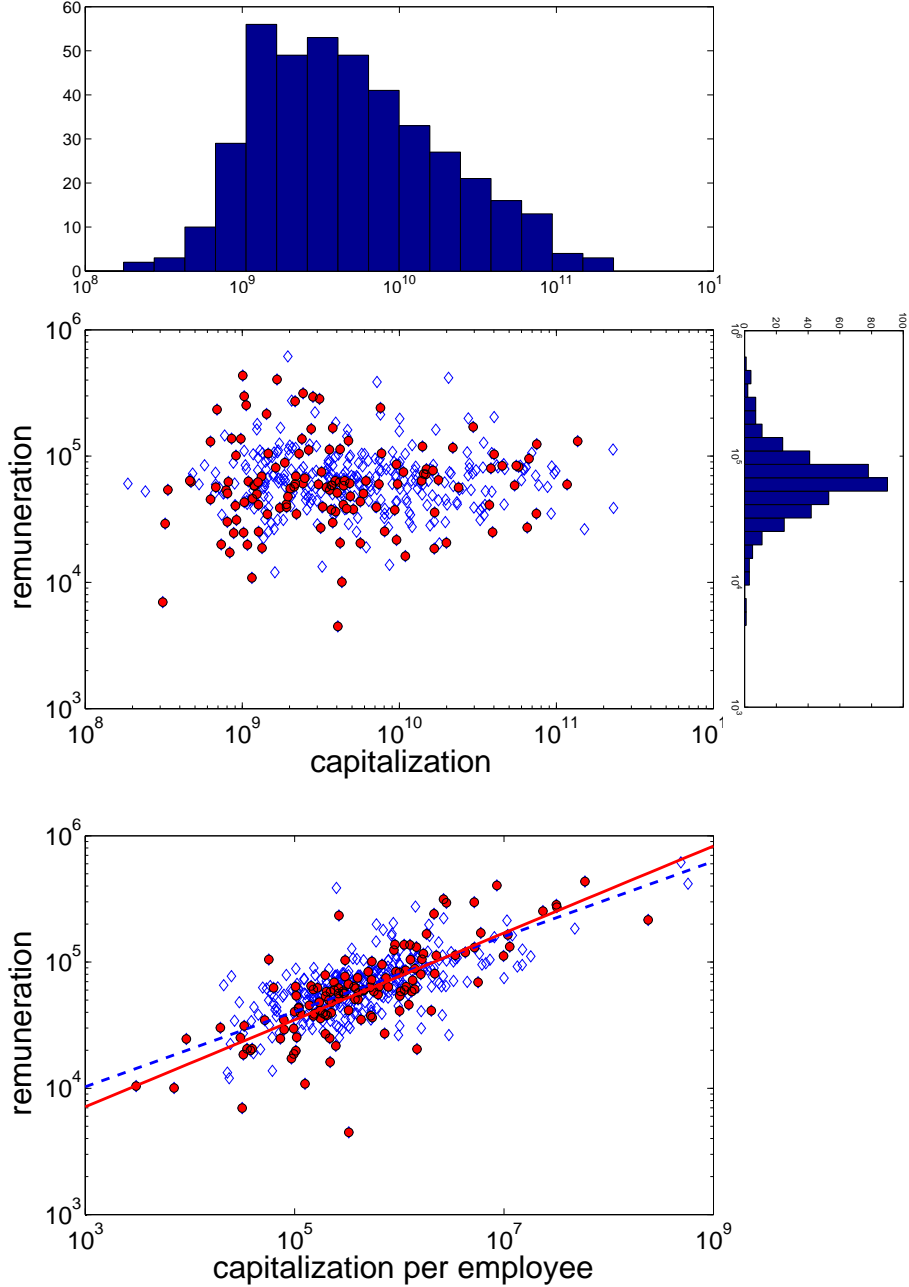}  
\end{center}
\caption{
Remuneration versus capitalization (top) and remuneration versus
capitalization per employee (bottom).  Full circles and empty diamonds
present large U.K. and European companies, respectively.  Both
quantities are shown in local currency and plotted on a logarithmic
scale to account for significant dispersion in capitalization and
remuneration.  Solid and dashed lines indicate the linear regression
between the logarithms of these quantities for the U.K. and European
universes, respectively (the respective slopes are 0.34 and 0.30, and
$R^2$ goodness of fit are 0.48 and 0.58, respectively).  Since the
records on remuneration and capitalization of each company in the
FACTSET database are updated at different moments of the year, data
were averaged over the period from 15/12/2014 to 30/07/2015.  Similar
results were obtained by taking the latest record for each company
(not shown).  Two subplots show the empirical distributions of
capitalization (top) and remuneration (right) among the biggest
European companies. }
\label{fig:capiremu}
\end{figure}

First, we inspect the empirical joint distribution of remuneration and
capitalization.  This inspection is important because a positive
size-wage effect has already been well documented in the economic
literature for microcapitalization firms \citep{Lallemand07}. The wage
gap due to firm size is approximately 35\% \citep{Oi99} because large
firms (but remaining in the microcapitalization category) demand a
higher quality of labor and set a higher performance standard that
must be supported by a compensating wage difference.  Note that the
magnitude and determinants of the employer-size wage premium vary
across industrialized countries.  Indeed, individual effects explain
approximately 90\% of inter-industry and firm-size wage differences in
France \citep{Abowd99}, while almost 50\% of the firm-size wage
differentials in Switzerland derive from a firm-size effect
\citep{Winter99}.  In the U.K., larger firms pay better because of
internal labor markets that reward effort and firm-specific capital
\citep{Belfieldk04}.  A visual inspection of Figure~\ref{fig:capiremu}
(top) suggests that there is almost no correlation between
remuneration and capitalization within the class of liquid stocks
(that excludes microcapitalization firms) and, in any case, residual
correlation is not significant.  As a consequence, a larger firm from
our sample does not necessarily pay its employees more.  This result
is consistent with the literature.

\begin{figure}
\begin{center}
\includegraphics[width=120mm]{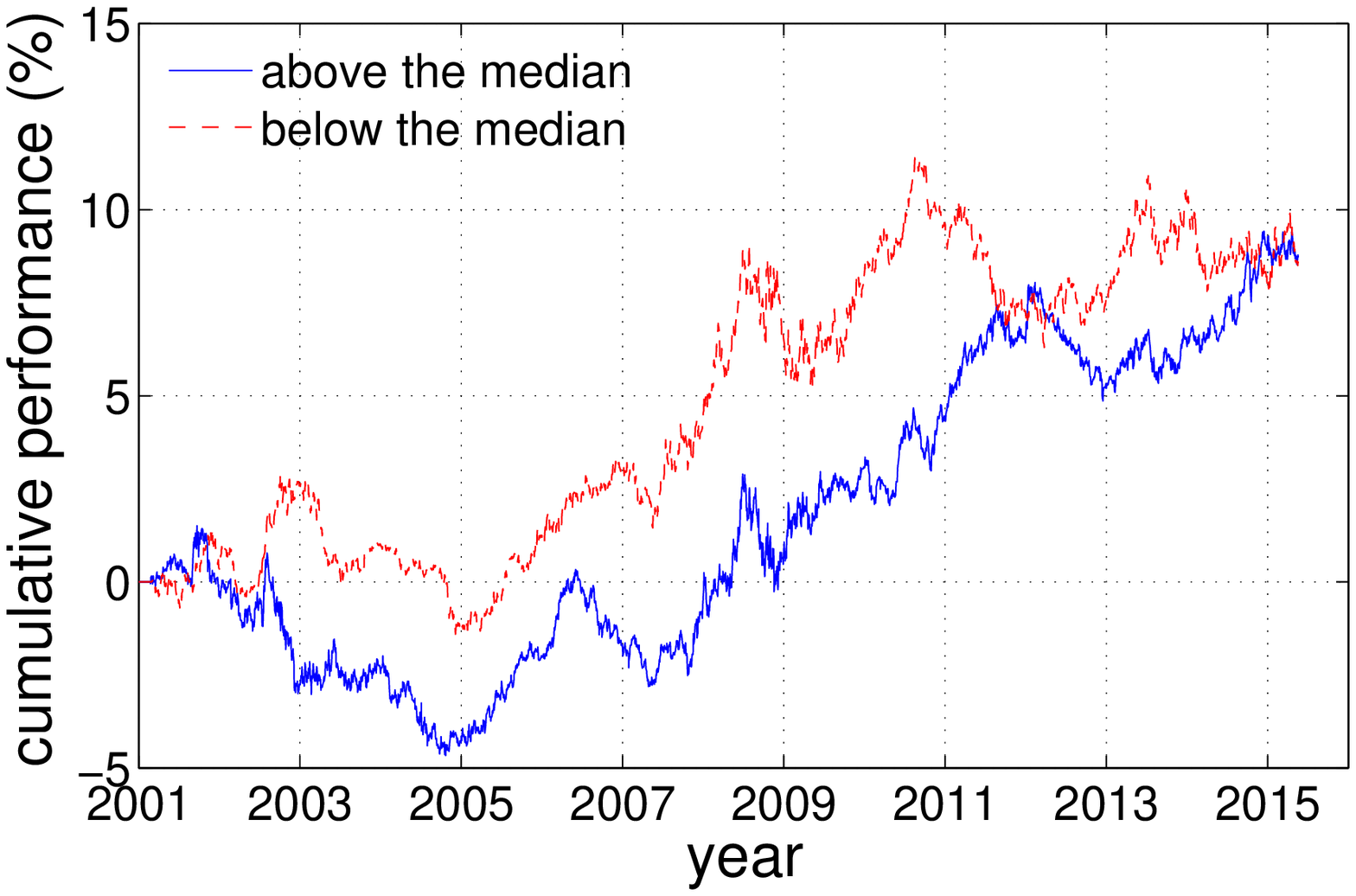} 
\end{center}
\caption{
Similar cumulative performance anomalies of two remuneration factors
for quantile Q1: one is constructed from stocks whose capitalization
exceeds the median size of our sample, and the other is constructed
from the remaining stocks.  The cumulative performance of both factors
after 15 years is approximately $9\%$, yielding an annualized
performance of $0.6\%$ (compared with $0.68\%$ in Table
\ref{tab:Sharpe}).  These curves are obtained for the European
universe (the results for the U.K. universe are similar and thus not
shown).  The annualized performance for the remuneration factor is
thus biased and cannot be fully explained by an unbiased random walk.}
\label{fig:Remu_capi}
\end{figure}

To confirm that the remuneration anomaly exists for different
capitalizations, we split our sample in two groups: the above-median
group of stocks whose capitalization exceeds the median size of our
sample, and the below-median group with the remaining stocks (we
recall that both groups exclude microcapitalization firms).  For each
group, we build its own remuneration factor.  Figure
\ref{fig:Remu_capi} shows that the cumulative performances of both
remuneration factors are statistically different from 0 and behave
similarly.  An apparent slight outperformance of the factor
constructed for the below-median group is not significant and can be
attributed to statistical fluctuations.


Further investigations on the size-wage effect compel us to explore
this relation per employee.  Figure \ref{fig:capiremu} (bottom)
reveals that {\bf remuneration is positively correlated to
capitalization per employee}, i.e., remuneration increases with the
amount of capitalization per employee.  One plausible explanation for
this phenomenon might be that reducing the number of employees (in
particular, underperforming employees) increases marginal
remuneration.  In summary, there is no correlation between
capitalization and remuneration for both universes of firms with
capitalization over 800 million euros.  Simultaneously, remuneration
increases with the amount of capitalization per employee -- as if the
cake had to be shared fewer times.

\subsection{Remuneration as a common risk}

\begin{figure}
\begin{center}
\includegraphics[width=120mm]{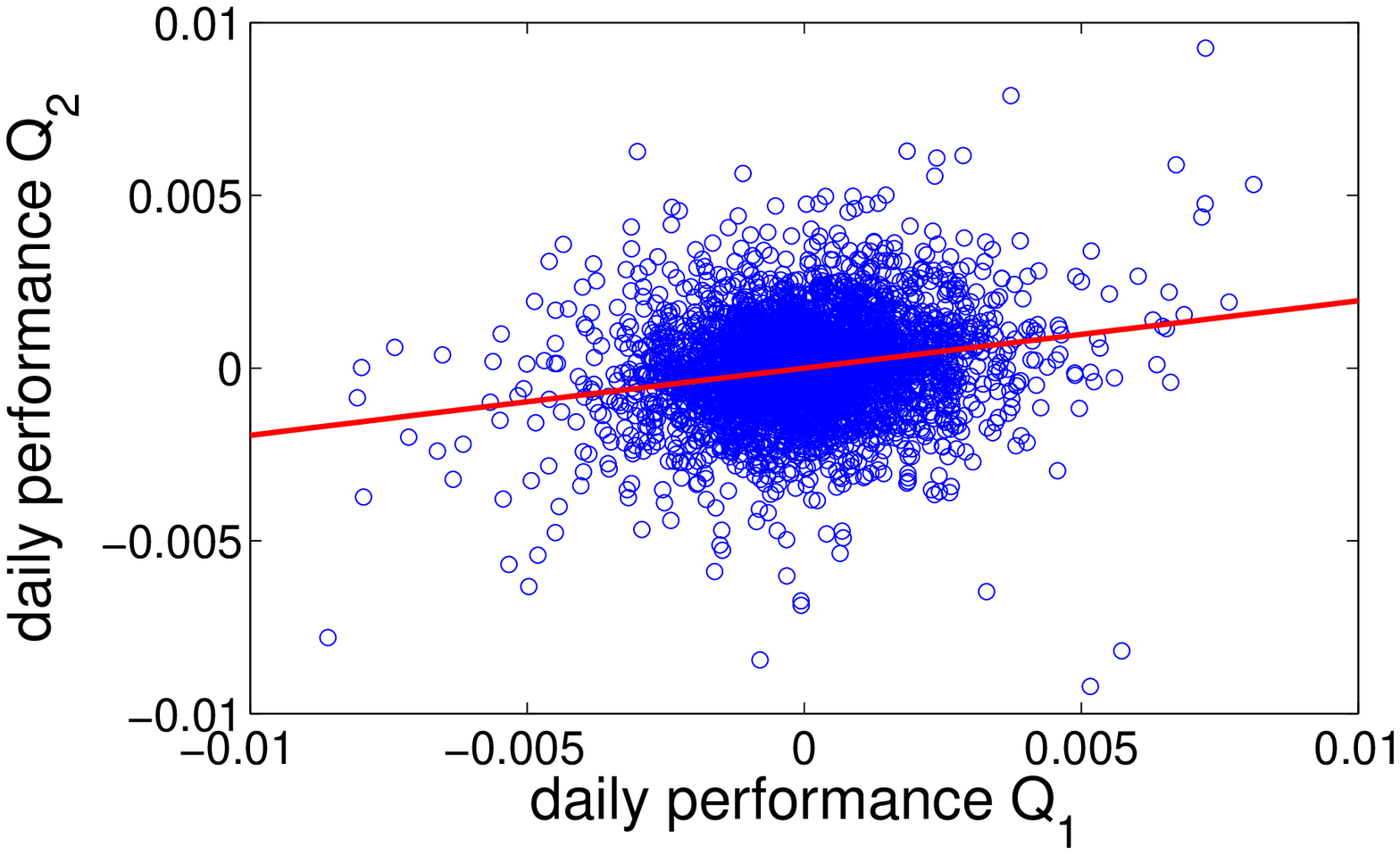}  
\includegraphics[width=120mm]{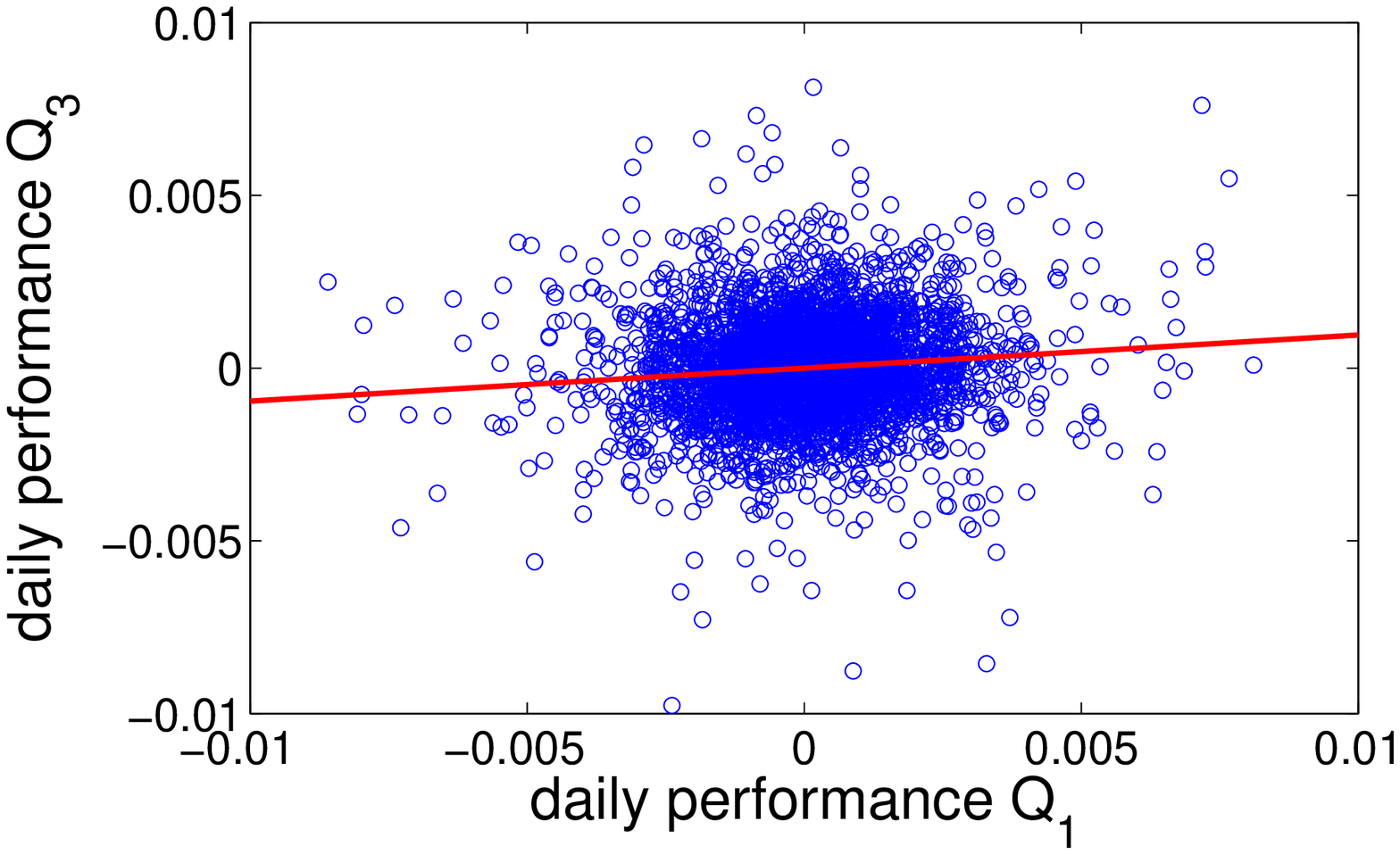}  
\end{center}
\caption{
{\bf (Top)} Correlation between the daily returns of the two
remuneration factors constructed on quantiles Q1 (0\%--15\% and
85\%--100\%) and Q2 (15\%--30\% and 70\%--85\%), which have no stocks
in common.  The daily returns of these factors are weakly correlated
but correlation is significant: the slope and its $95\%$-confidence
interval is $0.19 \pm 0.03$.  {\bf (Bottom)} For comparison, the
correlation between the daily returns of the remuneration factors Q1
and Q3 (30\%--45\% and 55\%--70\%) is shown, with the slope and its
$95\%$-confidence interval $0.10\pm 0.03$.  Both graphs were obtained
for the European universe.  Similar graphs for the U.K. universe yield
the slopes $0.23\pm 0.03$ and $0.02\pm 0.03$ for Q1-Q2 and Q1-Q3
correlations, respectively (graphs are not shown but are available
upon request).}
\label{fig:CumulBias}
\end{figure}

\begin{figure}
\begin{center}
\includegraphics[width=120mm]{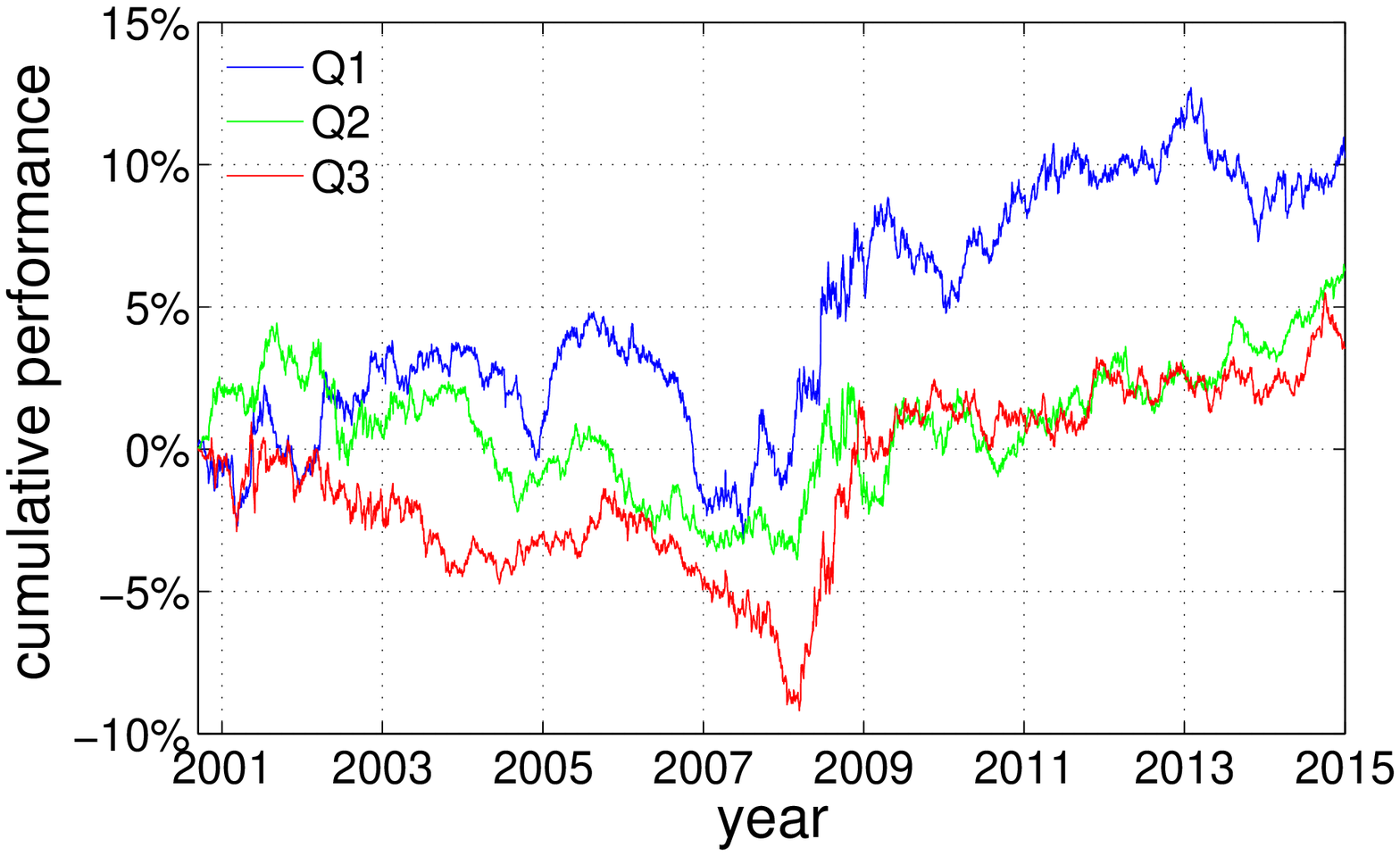}  
\end{center}
\caption{
The cumulative performance of the remuneration factor for the three
quantiles (Q1, Q2 and Q3) for the European universe (the graph for the
U.K. universe is similar and is available upon request).  Biases are
more pronounced for Q1 than for Q2 or Q3, which might be explained by
the possibility that stocks belonging to the extreme quantile are the
most sensitive to the remuneration anomaly.}
\label{fig:Remu}
\end{figure}

The motivation for building indicator-based factors relies on the
hypothesis that the stocks with close indicator values behave
similarly and thus share common risks. To verify this hypothesis, we
compare three realizations of the remuneration factor built on
different quantiles (Q1, Q2, and Q3), as described in
Section~\ref{sec:theory_factors}.  Figure~\ref{fig:CumulBias} shows
weak but highly significant correlation between the daily returns of
the remuneration factors from quantiles Q1 and Q2 (top) and Q1 and Q3
(bottom), notwithstanding that these factors have no stocks in common,
which is the indirect proof that {\bf the companies adopting similar
remuneration policies (e.g., paying their employees well) share a
common risk.}  The weak correlation can be explained by a rapid
decrease of the stock sensitivity to the remuneration factor with the
quantile: the correlation level of (Q1,Q3) is measured to be half that
of (Q1,Q2).  The common risk is of the same order of magnitude as the
residual risk, even for Q1.  In summary, only the stocks in the
extreme quantiles are the most sensitive to the remuneration factor.
This observation is also confirmed by {\bf the anomalies that are more
important for extreme quantiles}, as shown in Figure \ref{fig:Remu}.

\subsection{Factor correlation level as a proxy of the eigenvalues}

Ordering the factors based on their importance is central for the
asset pricing analysis.  As discussed in
Sec. \ref{sec:theory_factors}, the relevance of indicator-based
factors can be characterized using the factor correlation level (FCL)
defined by Eq. (\ref{eq:EV_q}).  If the factor weights were
approximately proportional to the elements of an eigenvector of the
correlation matrix, the FCL would be an estimator of the volatility of
this factor.  The factors with larger FCL would most likely have
greater impact on the portfolio returns for the same exposure.  In
general, the risk of a factor is proportional to the average
individual volatility multiplied by the FCL.  Thus, {\bf FCL can be
interpreted as an average correlation measurement between stocks
within the factor.}

\begin{figure}
\begin{center}
\includegraphics[width=120mm]{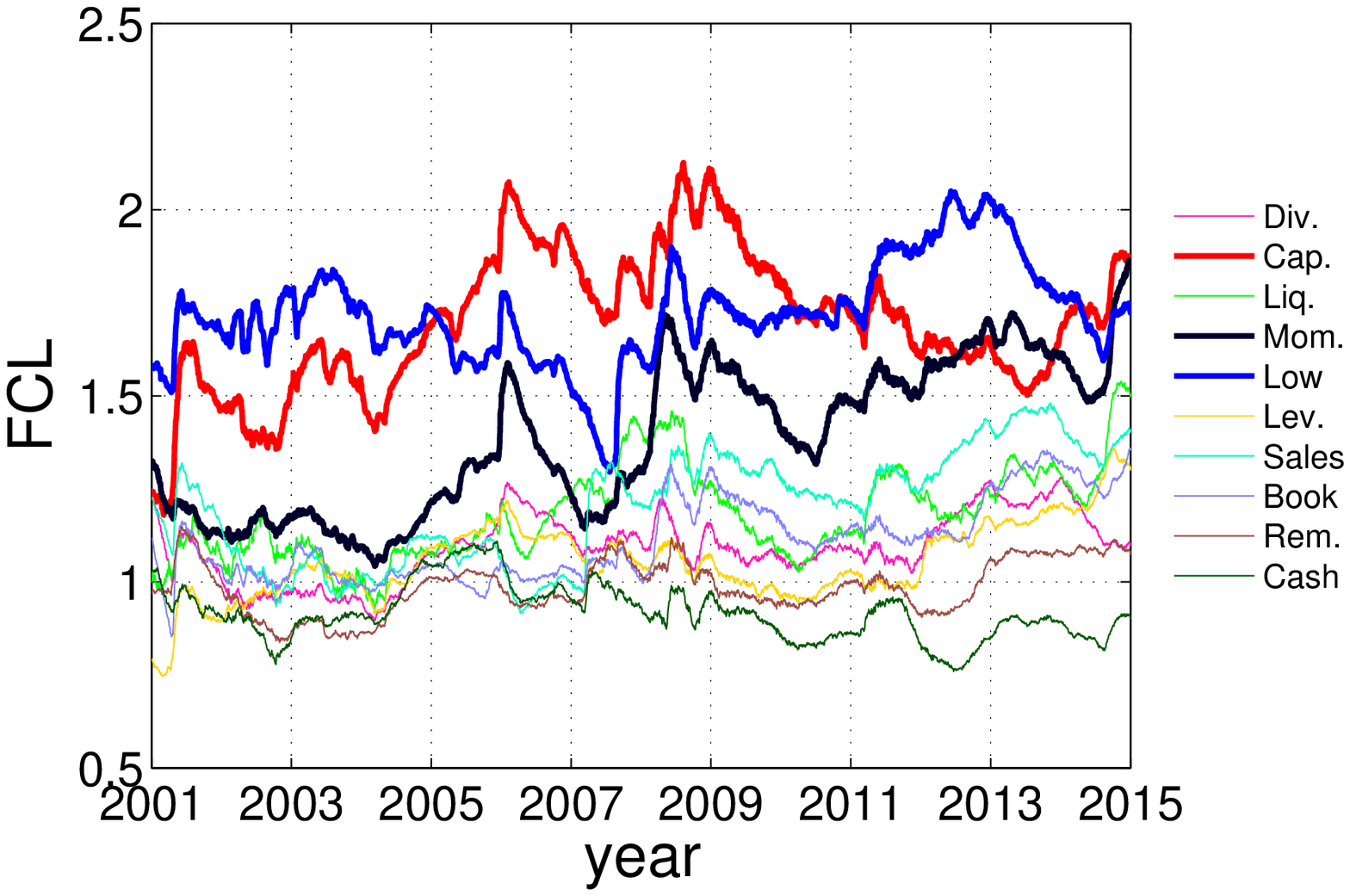} 
\includegraphics[width=120mm]{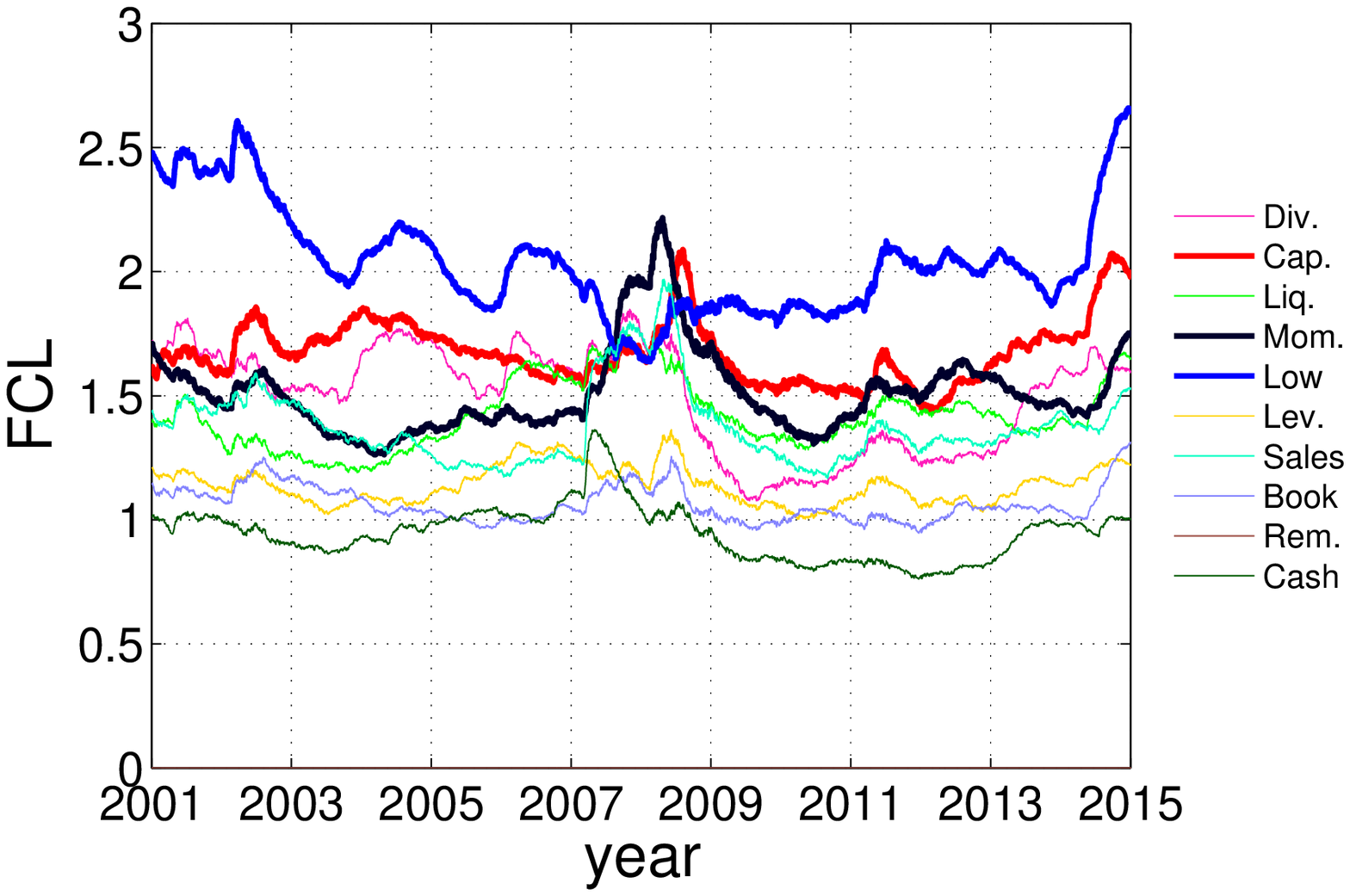} 
\end{center}
\caption{
Evolution of the factor correlation level (FCL) for ten factors
(quantile Q1): the European (top) and USA (bottom) universes (the
behavior for the U.K. universe is similar and available upon request).
In our interpretation, FCL is a measure of ``importance'' of factors
in asset pricing models.  Thick lines highlight the three major
factors: low-volatility, capitalization, and momentum.  The mean FCLs
averaged over 14 years are summarized in Table \ref{tab:FCL}.  All
FCLs are highly volatile, but this volatility is not linked to stock
market volatility.  In addition, we can see the jump- and cross-over
of FCLs.  During the 2007--2008 financial crisis, several FCLs
collapse for the U.S. universe.  Note that we could not construct the
remuneration factor for the U.S. universe because of lack of
systematic remuneration data for U.S. companies. }
\label{fig:VP}
\end{figure}

Using the daily returns of each factor and estimating the realized
volatility of each stock, we compute the FCL for each factor based on
Eq. (\ref{eq:EV_q}).  Figure \ref{fig:VP} shows the time evolution of
the FCLs for ten indicator-based factors defined in
Sec. \ref{sec:theory_factors}.  For comparison, we plot the FCLs for
the European and the U.S. universes (the FCLs for the U.K. universe
behave similarly and are thus not shown).  First, the FCLs exhibit
strong variations over time.  In particular, the FCLs of two factors
can cross each other, i.e., the ordering of the factors based on their
``importance'' can evolve over time.  For both universes, the
low-volatility factor appears as the most important, followed by
capitalization and momentum factors.  Other factors are smaller but
statistically significant.  Averaging the FCL over 15 years allows us
to order the factors according to their importance.  Table
\ref{tab:FCL} suggests {\bf the following order for the European
universe: low-volatility (1.73), capitalization (1.72), momentum
(1.41), sales-to-market (1.22), liquidity (1.19), book-to-market
(1.13), dividend (1.09), leverage (1.07), remuneration (0.99), and
cash (0.92).}  All these FCLs are higher than the noise level of
$0.78$ that we estimated by building a ``noise factor'' according to
an arbitrary non-financial indicator, such as an alphabetic order.
Even though the remuneration factor is relatively small, its magnitude
remains statistically relevant in comparison with other well-known
factors.  For example, the FCLs of the book-to-market, dividend,
leverage and cash factors are close to that of the remuneration
factor.  Their low values mean that these factors are not particularly
volatile and that the related common risks are low.  Conversely, the
low-volatility factor (excluded from the FF approach) has the highest
FCL and is thus identified as the first potential source of risk in a
portfolio, after market index and sectorial risks.  Notably, the
low-volatility factor is comparable to the capitalization factor and
greatly exceeds the book-to-market factor, the two ``major'' factors
identified in the
\citet{Fama93} model.

\begin{table}
\begin{center}
\begin{tabular}{| c | c | c | c | c | c | c | c | c | c | c || c |}  \hline
FCL    & Div. & Cap. & Liq. & Mom. & Low & Lev. & Sales & Book & Rem. & Cash & Market \\  \hline    
Europe & 1.09 & 1.72 & 1.19 & 1.41 & 1.73 & 1.07 & 1.22  & 1.13 & 0.99 & 0.92 & 10.41 \\  \hline     
U.K.   & 0.97 & 1.45 & 0.92 & 1.15 & 1.38 & 0.96 & 1.03  & 0.96 & 0.93 & 0.83 &  6.73 \\  \hline     
U.S.   & 1.49 & 1.73 & 1.49 & 1.62 & 2.10 & 1.15 & 1.41  & 1.12 & $-$  & 0.95 & 12.35 \\  \hline     
\end{tabular}
\end{center}
\caption{
~The mean value of the FCL for ten factors (quantile Q1) averaged over
the period from 10/08/2001 to 31/07/2015, for the European, U.K., and
U.S. universes.  According to these values, the main factors for asset
pricing are the low-volatility factor (excluded from the FF approach),
followed by the capitalization, and momentum factors.  We see that the
book-to-market and remuneration factors are of the same order of
magnitude such that the remuneration factor should have the same
importance in asset pricing models as the book-to-market factor.  We
also estimated the FCL of the market (last column).  The FCL of a
noise factor was estimated to be around $0.8$ for three universes
implying that all presented factors exceed noise.  Note that we could
not construct the remuneration factor for the U.S. universe because of
lack of systematic remuneration data for U.S. companies.}
\label{tab:FCL}
\end{table}

\subsection{Comparison with the principal component analysis}
\label{sec:PCA}

The principal component analysis (PCA), which is applied to
decorrelate time series, consists in forming the empirical correlation
matrix from daily stock returns and then finding its eigenvalues and
eigenvectors.  In practice, the number of stocks in a traded universe
(typically 500 - 1000) is often comparable to the number of available
historic returns per stock (for instance, 3612 daily returns in our
dataset), that makes this general method strongly sensible to noise,
as discussed in
\citep{Laloux99,Plerou99,Plerou02,Potters05,Wang11,Allez12}.  

\begin{figure}
\begin{center}
\includegraphics[width=120mm]{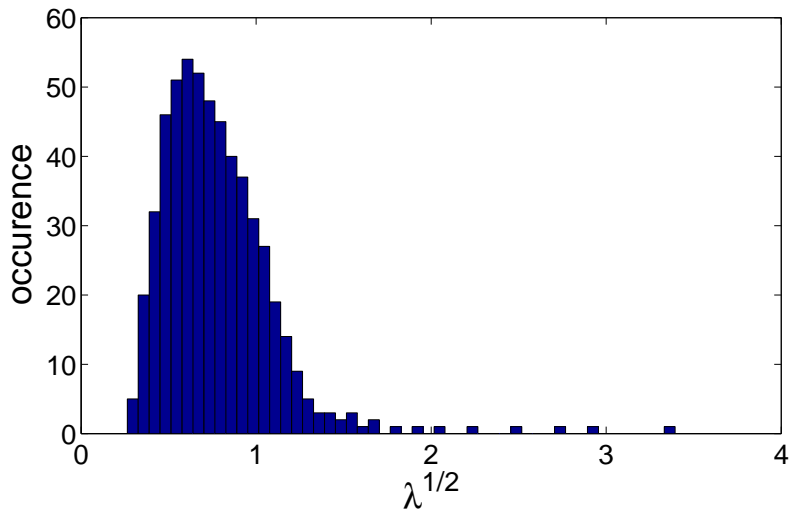}  
\end{center}
\caption{
Histogram of square roots of eigenvalues, $\lambda^{1/2}$, of the
empirical correlation matrix obtained from daily returns of 569 stocks
in the Europe universe over the period from 10/08/2001 to 31/07/2015.
The largest value, $\lambda^{1/2}_{\rm market} \approx 12.62$,
corresponding to the market mode, was excluded from the plot for a
better visualization of other values. }
\label{fig:PCA}
\end{figure}

In order to illustrate this limitation, we apply the PCA to the
European universe and compute numerically 569 eigenvalues.  Figure
\ref{fig:PCA} shows the histogram of square roots of the obtained
eigenvalues, i.e., how many eigenvalues are contained in successive
bins of size $0.0626$.  The largest value, $\lambda^{1/2}_{\rm market}
\approx 12.62$, corresponding to the market mode, was excluded from
the plot for a better visualization of other values.  One can identify
approximately ten well-separated single eigenvalues that are typically
attributed to market sectors.  In turn, the remaining part of
(smaller) eigenvalues lying close to each other and thus almost
indistinguishable, can be rationalized by using the random matrix
theory \citep{Laloux99}.  If the daily stock returns were distributed
as independent Gaussian variables (with mean zero and variance one),
the eigenvalues of the underlying empirical correlation matrix would
asymptotically be distributed according to the Marcenko-Pastur density
\begin{equation}
\rho(\lambda) = \frac{\sqrt{4q \lambda - (\lambda + q-1)^2}}{2\pi q \lambda} ,
\end{equation}
where $q = N/T$ is the ratio between the number of stocks, $N$, and
the number of daily returns per stock, $T$.  These eigenvalues lie
between two critical values, $\lambda_{\rm min} = (1-\sqrt{q})^2$ and
$\lambda_{\rm max} = (1+\sqrt{q})^2$.  As a consequence, the
eigenvalues obtained by diagonalizing the empirical correlation matrix
and lying below $\lambda_{\rm max}$ can be understood as statistical
uncertainty of the PCA.  In other words, the PCA cannot reliably
identify the factors with $\lambda < \lambda_{\rm max}$.  For our
European universe, $q = 569/3612$ so that $\sqrt{\lambda_{\rm max}}
\approx 1.4$ determines a theoretical threshold between larger,
significant eigenvalues, and smaller, noisy ones.

Comparing large values in Fig. \ref{fig:PCA} to the FCL from Table
\ref{tab:FCL}, we conclude that PCA might identify three major
factors: low-volatility (1.73), capitalization (1.72), and momentum
(1.41).  In turn, the other factors whose the FCL is smaller than the
PCA threshold $1.4$, would thus be understood as statistical
uncertainty in the PCA method.  {\bf The crucial advantage of our
method, in which factors are built from firm-based indicators while
market and sectorial correlations are eliminated by construction, is
the possibility to go beyond this PCA limit and to identify the
factors with smaller FCLs.}  Moreover, this identification can be
performed over time.

\subsection{Net investment as a proxy of the exposure to the low-volatility factor}

Building market-neutral portfolios requires nonzero net investment
when the portfolio is exposed to the low volatility anomaly.  This
anomaly is governed by the low-volatility factor, which is the most
influential factor (after market and sectors) according to our FCL
measurement (Table \ref{tab:FCL}), and unfortunately a residual
exposure to the low-volatility factor cannot be easily reduced.  As a
result, most factors can still be correlated to the low-volatility
factor.  Thus, when the average beta of long stocks in a factor is
significantly different from the average beta of short stocks, the
factor is also exposed to the low-volatility factor with a nonzero net
investment.  The net investment is defined as the difference between
long ($\omega_{i} >0$) and short ($\omega_{i} <0$) investments
normalized by total investment, i.e.,
\begin{equation}
\label{eq:Delta}
\Delta = \frac{\sum\nolimits_{i=1}^\N w_{i}}{\sum\nolimits_{i=1}^\N |w_{i}|} .
\end{equation}
By construction, $\Delta$ can vary between $-1$ and $1$ or,
equivalently, between $-100\%$ and $100\%$.

Replacing the individual sensitivities $\beta_{i}$ in the market
neutral relation (\ref{eq:beta_neutral}) by the averages $\langle
\beta_L\rangle$ and $\langle \beta_S \rangle$ for long and short
stocks, the net investment $\Delta$ from Eq. (\ref{eq:Delta}) can also
be expressed as
\begin{equation}
\label{eq:Delta_betas}
\Delta = \frac{\langle \beta_S \rangle - \langle \beta_L \rangle}{\langle \beta_S \rangle + \langle \beta_L \rangle}.
\end{equation}
When the average sensitivities for long and short stocks are similar,
net investment is close to $0$.  In turn, a net bias in $\Delta$
occurs when the average beta is different for long and short
stocks. $\Delta$ is a proxy of the exposure to the low-volatility
factor that is more reactive and more precise than the estimation
obtained through the usual regression of returns.

The bias in the long and short betas in Eq. (\ref{eq:Delta_betas}) may
also be related to the sensitivity to the market (i.e., to the stock
index) of a factor built with the FF approach (i.e., neutral in
nominal but not in beta):
\begin{equation}
\beta_{FF} = \langle \beta_L \rangle - \langle \beta_S \rangle = - 2\langle \beta\rangle \Delta, 
\end{equation}
where $\langle\beta\rangle = \frac12(\langle \beta_S \rangle + \langle
\beta_L \rangle)$ is the average beta of the universe that we
estimated as $\langle\beta\rangle \approx 0.65$ for the period from
2001 to 2015.  The net investment $\Delta$ can also be related to the
sensitivity of any beta neutral portfolio or factor (both in the FF
approach and in our methodology) to the low-volatility factor (the
most influential factor, according to the FCL).

\begin{figure}
\begin{center}
\includegraphics[width=120mm]{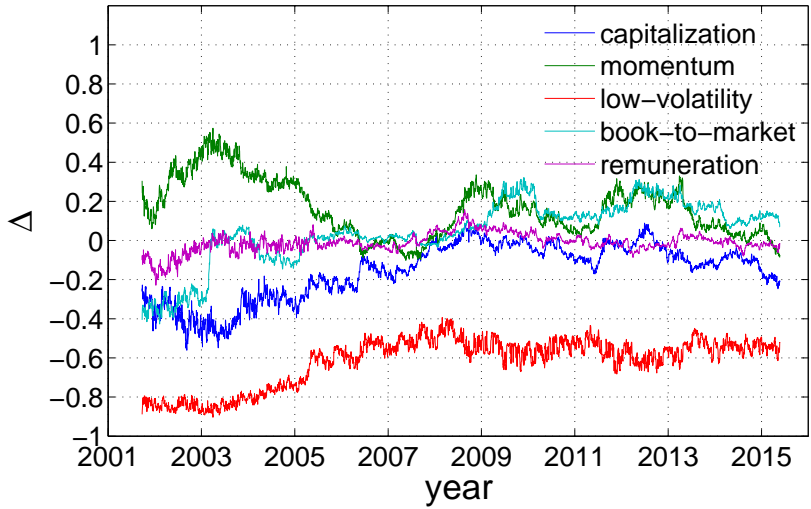} 
\end{center}
\caption{
Evolution of the net investment $\Delta$ for five indicator-based
factors for the European universe: capitalization, momentum,
low-volatility, book-to-market, and remuneration (the results for the
U.K. universe are not shown but are available upon request).  We
recall that $\Delta$ is a proxy of the exposure to the low-volatility
factor.  The remuneration $\Delta$ is around zero, and the factor
therefore has no correlation with the low-volatility factor.  Other
factors seem to be more exposed to the low-volatility factor.}
\label{fig:NetInv}
\end{figure}

Figure \ref{fig:NetInv} shows that the low-volatility factor has the
most important short investment (negative values of $\Delta$ ranging
between $-80\%$ and $-60\%$), although its sensitivity to the market
was maintained at 0.  Other factors also have a bias in $\Delta$,
including the capitalization and the momentum factors, in particular.
In the FF approach, these factors would therefore also have a
significant sensitivity to the market.  In particular, the
low-volatility factor built with the FF approach would be strongly
correlated to the market.  Moreover, $\Delta$ indicates that most
factors have a residual correlation with the low-volatility factor
that remains uncorrected by our method.  Since 2003, the $\Delta$ of
the book-to-market factor (one of the major anomalies investigated by
Fama and French) has shrunk, and {\bf the related book-to-market
anomaly has almost disappeared} (see Table \ref{tab:Sharpe}).
Finally, the remuneration factor shows nearly zero net investment,
i.e., it remains uncorrelated with the low-volatility factor.

\subsection{Other inter-factor correlations}

Correlations between factors matter as long as one needs uncorrelated
portfolios for asset pricing purposes.  The indicator-based factors
were introduced to build as many uncorrelated portfolios as
possible. At the same time, such an explicit construction does not
guarantee to yield truly uncorrelated combinations, such as the
eigenvectors of the covariance (or correlation) matrix.  Moreover,
some indicators may capture the same economic or financial features of
the company and may thus be correlated; in other words, different
factors may approximate the same eigenvector and thus be highly
correlated.  In particular, adding new indicator-based factors does
not necessarily help to capture new features and may thus be
redundant.  The choice of the ten indicator-based factors studied in
this paper is judged as sufficient with respect to the trade-off
between capturing information and remaining uncorrelated.
Table~\ref{table:multireg} presents the correlation coefficients
between ten indicator-based factors estimated from their
volatility-normalized daily returns.  Clearly, many indicator-based
factors remain correlated.  If the same estimation was applied to ten
independent Gaussian vectors of the same length ($m = 3612$ elements),
the standard deviation of the estimated correlation coefficients would
be $1/\sqrt{m} \approx 0.0166$.  In other words, the presented
correlations between the indicator-based factors are highly
significant.

The remuneration factor exhibits correlations with some other factors,
and the most significant of these include the following: the
sales-to-market ($-0.38$), dividend ($-0.23$), and momentum ($0.20$)
factors.  These correlations can be explained as follows.  First, {\bf
the companies with low sales-to-market ratios have a high margin and
thus the ability to pay their employees well} (strong negative
correlation $-0.38$).  The direct link between a firm's margin and
wage is well documented in the labor economics literature.  More
precisely, there is a relation between margin and labor cost. For
instance, a study by the European Central Bank (ECB) and the
Organization for Economic Co-operation and Development (OECD) reveals
that larger firms make more extensive use of margin for labor
cost-cutting strategies, i.e., firms choose to reduce benefits as a
cost-cutting strategy \citep{Babecky12}.  In addition, the positive
relation between firm size and the use of cost-cutting strategies that
is monotonically increasing and highly significant, is uncovered.
Second, {\bf the companies that pay high dividends to shareholders
tend to remunerate their employees less}, yielding a negative
correlation of $-0.23$, which is a direct representation of
profit-sharing within firms.  Indeed, dividend payments are charged on
the profits of the business after all salaries and benefits expenses
are paid out.  Although this result appears intuitive, it remains
important as it reveals the level of correlation between both
quantities.  The labor economics literature and the corporate finance
literature are not very well documented on this particular issue.
Finally, {\bf companies that perform well and show strong momentum can
offer higher remuneration to their employees or, alternatively, the
higher remuneration stimulates employees to work better and to imbue
the company with momentum} (positive correlation $0.20$).  This is a
central and very important result of our research because it
highlights the positive relation between pay and performance.  The
rationale behind this result is discussed in Section
\ref{sec:discussion}.

\begin{table}
\begin{center}
\begin{tabular}{| c | c c c c c c c c c c |}  \hline
      &   Div. &  Cap.    & Liq.    &   Mom.   & Low    & Lev.     & Sales.    &  Book.   & Rem.    &  Cash   \\  \hline
Div.  &           & 0.10  & 0.14  &{\bf -0.33}& 0.02 & 0.29  & 0.26      & 0.18      & -0.23     & 0.14  \\ 
Cap.  & 0.10      &       & 0.08  & 0.10      & 0.13 & 0.21  & -0.20     & -0.06     & 0.05      & -0.01 \\ 
Liq.  & 0.14      & 0.08  &       & -0.21     & 0.20 & 0.09  & 0.05      & 0.05      & -0.06     & 0.05  \\ 
Mom.  &{\bf -0.33}& 0.10  & -0.21 &           & -0.18& -0.24 & -0.25     &{\bf -0.36}& 0.20      & -0.04 \\ 
Low   & 0.02      & 0.13  & 0.20  & -0.18     &      & 0.02  & 0.01      & 0.07      & -0.03     & 0.05  \\ 
Lev.  & 0.29      & 0.21  & 0.09  & -0.24     & 0.02 &       & 0.23      & 0.11      & -0.17     & -0.02 \\ 
Sales.  & 0.26      & -0.20 & 0.05  & -0.25     & 0.01 & 0.23  &           &{\bf 0.31} &{\bf -0.38}& 0.23  \\ 
Book.  & 0.18      & -0.06 & 0.05  &{\bf -0.36}& 0.07 & 0.11  &{\bf 0.31} &           & -0.13     & 0.05  \\ 
\rowcolor[gray]{0.9}
Rem.  & -0.23     & 0.05  & -0.06 &  0.20     & -0.03& -0.17 &{\bf -0.38}& -0.13     &           & -0.11 \\ 
Cash & 0.14      & -0.01 & 0.05  & -0.04     & 0.05 & -0.02 & 0.23      & 0.05      & -0.11     &       \\  \hline
\end{tabular}
\end{center}
\caption{
~Correlation coefficients between 10 indicator-based factors for the
U.K. companies: Dividend (1), capitalization (2), liquidity (3),
momentum (4), low-volatility (5), leverage (6), sales-to-market (7),
book-to-market (8), remuneration (9), and cash (10).  These
coefficients were estimated from daily returns of these factors over
the period from 23/02/2001 to 27/07/2015.  Daily returns of each
factor were normalized by their volatility averaged over 20 days to
reduce the effects of heteroskedasticity.  Similar correlation
coefficients were obtained for the European companies (available upon
request).}
\label{table:multireg}
\end{table}

\begin{figure}
\begin{center}
\includegraphics[width=120mm]{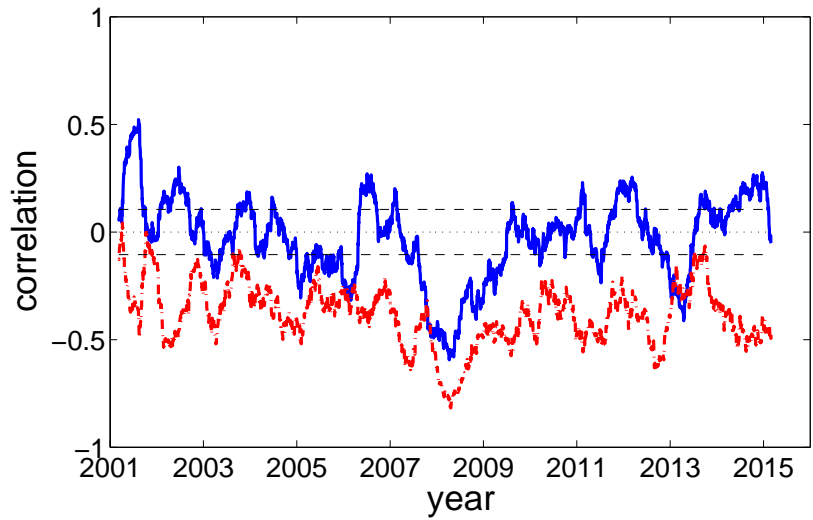}  
\end{center}
\caption{
Correlation coefficients between daily returns of the remuneration
factor and of the low-volatility factor (solid line) or the
sales-to-market factor (dashed line) for the largest U.K. companies.
The coefficients were computed over a sliding window of 90 days.
Prior to computation, the daily returns were renormalized by their
average volatility over the previous 20 days.  The mean values over 15
years are $-0.03$ and $-0.38$ (see Table \ref{table:multireg}),
respectively.  Horizontal dashed lines show the standard deviation,
$0.105$, of the same estimator applied to two independent Gaussian
samples.  Similar results were obtained for the European universe
(available upon request). }
\label{fig:corr_time}
\end{figure}

It is worth emphasizing that these correlations between factors are
not static (as presented in Table~\ref{table:multireg} by averaging
over 15 years) but evolve over time.  For example,
Fig. \ref{fig:corr_time} shows the evolution of two correlation
coefficients between volatility-normalized daily returns of
remuneration, low-volatility, and sales-to-market factors.  The
correlation between the remuneration and low-volatility factors
remains close to zero, with eventual deviations beyond the Gaussian
significance range (e.g., during the subprime and financial crises in
2007-2009).  These two factors can be considered uncorrelated.  In
turn, the negative correlation between the remuneration and
sales-to-market factors always remains beyond the Gaussian
significance range.

\subsection{The anomaly of the remuneration factor and its interpretation}

Table \ref{tab:Sharpe} compares the remuneration anomaly with other
factors in terms of the annualized bias (the annualized cumulative
return between the last and the first observation days), the Sharpe
ratio (the annualized bias normalized by annualized volatility), and
t-statistics (the Sharpe ratio multiplied by the square root of the
total duration in years).  In particular, the t-statistic allows one
to reject the null hypothesis of no bias at the 90\% confidence level.

The bias reveals the level of overperformance due to a particular
factor. We observe a significant bias for the dominant capitalization
and low-volatility factors, which have been previously documented.
The anomaly of the book-to-market factor seems to have disappeared
(see Table \ref{tab:Sharpe}).  In fact, the Sharpe ratio that we
estimated to be 0.49 for the period from 1926 to 2008 in the
U.S.\footnote{Based on the publicly available data from Fama and
French,
\url{http://mba.tuck.dartmouth.edu/pages/faculty/ken.french/data_library.html}},
became much smaller in recent years (and even changed the sign for the
European universe, becoming $-0.08$).  We suspect that this result can
be explained by the change in its exposition to the low-volatility
factor.  The momentum factor has also changed direction.

The remuneration factor appears as the sixth most important anomaly in
the U.K. market, and the eighth most important anomaly in the European
market.  {\bf A bias of $1.21\%$ means that companies that pay better
should overperform their less paying competitors by $2\times 1.21\%$.}
The prefactor $2$ appears if we assume that 50\% is invested in high
remuneration and $50\%$ in low remuneration (i.e., there is no
exposure to the low-volatility factor and volatility is nearly
homogeneously distributed).  This is one of the most important results
in this paper, as it shows that {\bf a market neutral investment style
arbitrage strategy based on the remuneration anomaly is likely to
deliver positive returns.}  Next, assuming that the bias in the
remuneration factor consists of an intrinsic bias and contributions
from biases of other factors due to inter-factor correlations, the
relative impacts of these biases can be estimated by multiplying them
by the correlation coefficients in the $9^{\rm th}$ line of Table
\ref{table:multireg}.  These relative impacts are summarized in the
last line of Table \ref{tab:Sharpe}.  Since most contributions from
other factors are negative, it might be surmised that the intrinsic
remuneration bias is even higher than $1.21\%$ (estimated to be around
2.85\%) but that its value is reduced due to correlations with other
factors.  If we were able to build a remuneration factor fully
decorrelated from other factors, we would have obtained most likely a
t-statistic above 3 (around $3.29$, see Table \ref{tab:Sharpe}) that
fulfills the requirements formulated by \citet{Harvey15a}.  Note also
that there is no selection bias in our study (we have not analyzed all
the different possibilities to finally retain the remuneration
factor), such that the condition requiring a t-statistic greater than
3 when taking into account the number of possible anomaly candidates
is not applicable.  In any event, the observed bias of $1.21\%$ cannot
simply be explained by the biases of other factors.  The Sharpe ratio
of $0.37$ indicates that a horizon of $1/0.37 \approx 2.7$ years is
required for the anomaly to be captured and to have a positive return
with a likelihood of 84\%.  From an asset management point of view, it
suggests the recommended time horizon to take profits based on this
market anomaly.

\begin{table}
\begin{center}
\begin{tabular}{| c | c | c c c c c c c c >{\columncolor[gray]{.9}} c c|} \hline
&  &  Div. &  Cap.    & Liq.    &   Mom.   & Low    & Lev.     & Sales.    &  Book.   & Rem.    &  Cash   \\  \hline
\multirow{3}{3mm}[-1mm]{\begin{turn}{90}Europe\end{turn}}
&  Bias, \% & 2.39 & -5.72 & -0.95 & -1.60 & -4.15 & -1.95 & 0.08 & -0.23 & 0.68 & 1.66 \\   
& Sharpe    & 0.80 & -1.69 & -0.41 & -0.42 & -1.46 & -0.74 & 0.03 & -0.08 & 0.25 & 0.65 \\   
& t-stat    & 3.04 & -6.42 & -1.57 & -1.59 & -5.57 & -2.81 & 0.11 & -0.30 & 0.97 & 2.46 \\ \hline  
\multirow{3}{3mm}[-1mm]{\begin{turn}{90}U.K.\end{turn}}
&  Bias, \% & 2.12 & -4.29 & -0.11 & -2.81 & -3.81 & -1.01 & 0.92 &  0.34 & 1.21 & 2.60 \\   
& Sharpe    & 0.65 & -1.38 & -0.05 & -0.71 & -1.25 & -0.35 & 0.31 &  0.11 & 0.37 & 0.92 \\   
& t-stat    & 2.48 & -5.24 & -0.18 & -2.71 & -4.77 & -1.34 & 1.16 &  0.40 & 1.40 & 3.51 \\ \hline  
& Impact, \%  &-0.49 & -0.21 &  0.01 & -0.56 &  0.11 &  0.17 &-0.35 & -0.04 & 1.21 & -0.29 \\ \hline  
\end{tabular}
\end{center}
\caption{
~The annualized bias (the annualized cumulative return between the
last and the first observation days, as a percentage), the Sharpe
ratio (annualized bias normalized by annualized volatility), and the
t-statistic (the Sharpe ratio multiplied by the square root of the
total duration in years, i.e., by $\sqrt{14.5} \simeq 3.81$) for the
following 10 indicator-based factors (quantile Q1): dividend (1),
capitalization (2), liquidity (3), momentum (4), low-volatility (5),
leverage (6), sales-to-market (7), book-to-market (8), remuneration
(9), and cash (10).  These quantities are estimated for the period
from January 2001 to July 2015, for the largest European companies
(top lines) and for the largest U.K. companies (bottom lines).  The
last line shows the relative impacts of the biases of various factors
on the remuneration bias (1.21) for the U.K. companies.  These impacts
are obtained by multiplying the biases in the fourth line by the
correlation coefficients from the $9^{\rm th}$ line of Table
\ref{table:multireg}.  The annualized bias for the remuneration factor
in the U.K. universe is 1.21\% with a t-statistic of 1.21.  Moreover,
if we subtract all the impacts from remuneration's annualized bias, we
obtain an intrinsic remuneration bias of 2.85\%.  Therefore, we would
have a t-statistic of approximately $2.85 \times 1.40/1.21 = 3.29$
that would fulfill the requirements formulated by \citet{Harvey15a}.}
\label{tab:Sharpe}
\end{table}


\subsection{The rationale behind the remuneration anomaly}

In a survey paper, \citet{Yellen84} poses the question of why firms do
not cut wages in an economy characterized by involuntary unemployment?
Indeed, unemployed workers would prefer to work at the real wage
rather than being unemployed, but firms will not hire them at a lower
wage simply because any reduction in wage would lower employee
productivity.  This is Yellen's most-cited paper, and it stipulates
that the amount of effort that employees put into their job depends on
the difference between the wage they are getting paid and what they
perceive as a ``fair wage''.  The bigger the difference, the less hard
they tend to work, which highlights the idea that paying employees
more than the market clearing wage may boost productivity and ends up
being worthwhile for the employer.  Paradoxically, cutting wages may
end up raising labor costs since it will negatively affect
productivity \citep{Stiglitz81}. Hence, productivity is the main
argument, which is confirmed by other theoretical papers that consider
employees to be more productive in larger firms and thus explain why
they demand higher wages \citep{Idson99}.  The other arguments are as
follows.  Given job contract incompleteness, not all duties of an
employee can be specified in advance.  For this reason, monitoring is
a central instrument to control production costs \citep{Alchian72}.
Unfortunately, monitoring is too costly and sometimes inaccurate due
to measurement error.  Instead of having costly and imperfect
monitoring, firms can offer higher wages to their employees to create
an incentive for the employee not to lose their high wage by being
fired \citep{Shapiro84}.  In this context, paying a wage in excess of
the market clearing wage can be seen as an efficient way to prevent
employees from shirking.  The attractiveness of wages to skillful
workers also contributes to reduce their turnover.  Moreover, raising
wages partly eliminates job demands from less performing candidates
who would fear competing with overperforming candidates. This adverse
selection is a subtle support for the fair wage hypothesis because
paying fair wages will attract only the more skillful workers and
deter lemons and will thus help avoid costly monitoring devices in the
recruitment processes.  In summary, the motivation for the fair
wage-effort hypothesis is a simple observation of human nature arguing
that employees who receive less than what they perceive to be a fair
wage will not work as hard as a consequence.  In the very same vein,
\citet{Akerlof90} set up a model of unemployment in which ``people
work less hard if they are paid less than they deserve, but not harder
if they receive more than they deserve''.  The model puts in equation
the fair wage-effort hypothesis to represent the idea that a poorly
paid employee may be keen on taking its revenge on its employer.

\section{Discussion}
\label{sec:discussion}

\subsection{Fama and French approach}

\citet{Fama93,Fama15} use time series of 25 portfolios, each
portfolio built with similar capitalization and book-to-market
stocks. They regress the monthly performance $R_i(t)$ of each
portfolio $i$ on the returns $f_j(t)$ of different factors $j$:
\begin{equation*}
R_i(t) = a_i + \sum\limits_j b_{i,j} f_j(t) + \varepsilon_i(t),
\end{equation*}
where $a_i$ and $\varepsilon_i(t)$ are portfolio-specific intercept
and noise, and $b_{i,j}$ is the estimated sensitivity of the $i$-th
portfolio to the $j$-th factor.

If the remuneration factor had to be investigated using the FF
approach, how could one proceed?  Five different portfolios might be
built with stocks sorted according to remuneration and then at least
three major factors might be used: the market index, capitalization,
and book-to-market factors (the factor returns, $f_j(t)$, would be
estimated through the performance of the long-short portfolio, e.g.,
buying the high capitalization and shorting the low capitalization, or
buying the high book-to-market and shorting the low book-to-market).
The intercept, $a_i$, for the 5 different portfolios might be measured
with their t-statistics to assess whether the remuneration is an
anomaly.  One might also measure the $a_{\rm high} - a_{\rm low}$ and
its t-statistics, as in Table 2 by \citet{Fama08}.  Finally, the
remuneration factor might be added to the regression panel and the
$R^2$ for every portfolio might be measured to quantify how well the
data fit the statistical model and how well the common factors explain
the price returns.

Instead, we simply measure the average returns of the HML portfolio
(see Table \ref{tab:Sharpe}) built to be beta-neutral without any
regression, as we construct our remuneration factor as uncorrelated to
the main factors.  That should be close to the $\frac12(a_{\rm high} -
a_{\rm low})$ of the FF approach, or close to the average return of
the HML portfolio built to be delta-neutral (see Table I from
\citet{Fama15b}).  This is due to the fact that the remuneration
factor is not exposed to the market index, low-volatility and
book-to-market factors.  However, the FF approach would not account
for the fact that remuneration depends on sectors (see Table
\ref{tab:book-to-market}).  Using the volatility of the portfolio, we
can also measure the t-statistics to learn whether the anomaly is
statistically significant, and we measure the FCL to quantify how well
the common factors explain the price returns.

In Appendix \ref{sec:FF} we compare the FF approach to our
methodology.  In particular, we show that sectorial constraint and
beta-neutral property were the two key advantages of our factors
construction: without them, {\bf the FF approach applied to the same
period, would give insignificant results for the remuneration factor}
(we recall that most Fama and French data begin from 1963, which leads
to greater t-statistics).

\begin{table}
\begin{tabular}{| l | r | r |}  \hline
Sectors          & Median book-to-market & Median remuneration (in euros) \\  \hline
Consumer discretionary & 0.31443 &  22 859.96 \\
Consumer staple & 0.24681 &  39 416.51 \\
Energy           & 0.81440 & 137 625.91 \\
Financial        & 0.87972 & 126 498.10 \\
Health           & 0.24442 &  51 452.06 \\
Industrial       & 0.32765 &  58 626.27 \\
IT               & 0.19867 &  77 854.94 \\
Material         & 0.55733 &  32 516.14 \\
Telecom          & 0.39122 &  66 283.21 \\
Utilities        & 0.32572 &  47 014.69 \\  \hline
\end{tabular}
\caption{
~Sectorial variations of the median of the book-to-market and of the
remuneration (in euros) for the U.K. universe in 2014.  Both
book-to-market value and remuneration vary substantially across
different sectors.}
\label{tab:book-to-market}
\end{table}

\subsection{Advantages and limitations of the methodology}

Our methodology has several advantages over the FF approach:
\begin{enumerate}
\item
The estimated FCL quantifying the relevance of the factor does not
depend on the number of considered factors, in contrast to the $R^2$
argument of the FF approach (e.g., see Table 6 in
\citet{Fama93}).  Thus, one can select the most important factors
(e.g., stock index, low-volatility, capitalization, liquidity, and
momentum factors) in asset pricing models.

\item
The sensitivities of the different common risk factors to the market
(i.e., to the stock index) are maintained at zero even for the
low-volatility factor, which is an important feature because the
market mode may have a hundred times greater impact on portfolio
returns than other factors.

\item
The factors are constructed to be sector neutral, which allows one to
better identify their impacts on price variations, which is important
because intra-sector correlations are typically more important than
within-factor correlations.  Notably, the book-to-market factor of FF
approach also captures sectorial risk, as the firms are not priced in
the same way from one sector to another (see Table
\ref{tab:book-to-market}).  In particular, the remuneration is very
different from one sector to another.

\item
Weights ($w_i$) of the stocks that are close in capitalization (or in
book-to-market, or in remuneration, etc., depending on the factor) are
of the same order of magnitude that reduces the specific risk of the
factor.

\item
Maintaining factors beta-neutral at any time reduces the noise of
factors, even those that are not supposed to be correlated to the
stock index.  In fact, we will show in Appendix \ref{sec:FF} that in
the case of factors uncorrelated to the stock index, the beta-neutral
constraint reduces the volatility of the factor by $1.2\%$ on an
annualized basis.

\item
Our method enables the inclusion of the low-volatility factor into the
cross-section of average returns (in contrast to the FF approach)
without any multiregression model.  The low-volatility and
capitalization factors were found to provide the largest anomaly (see
Table \ref{tab:Sharpe}).  In addition, the low-volatility factor was
also identified as the major contribution to risk, according to our
measurement (see Fig.~\ref{fig:VP}).  Surprisingly, the capitalization
factor, which had previously been considered as the most important,
now occupies the second position.  Moreover, the book-to-market factor
identified by \citet{Fama93} as important, has eventually become a
minor factor (and is just slightly more important than the
remuneration factor) after having eliminated the sectoral and market
modes.

\end{enumerate}

The main limitations to our methodology are related to the methodology
itself.  Indeed, although introducing indicator-based factors and
their relevance assessments through the FCL were inspired by
eigenbasis, this construction does not pretend to yield true
eigenvectors and eigenvalues of the covariance (or correlation)
matrix.  In particular, correlations observed between several factors
(e.g., the remuneration and sales-to-market factors) indicate that the
decorrelation performed is not perfect.  Although the construction of
factors can be further refined to make them less correlated (e.g., by
splitting the stocks into smaller groups than supersectors), it is
difficult to quantitatively assess the quality of such improvements.

\section{Conclusion}
\label{sec:conclusion}

We identify a new anomaly in asset pricing that is statistically
significant and economically relevant.  It is linked to remuneration:
the more a company pays for salaries and benefits expenses per
employee, the better its stock performs.  We show that remuneration is
a common risk factor although its magnitude appears relatively small
compared with dominant factors such as low-volatility or
capitalization.  It also appears that only the companies that belong
to extreme quantiles are sensitive to the remuneration factor.  To
validate the abnormal performance associated with the remuneration
factor, we check that performance is not explained by other major
factors such as low-volatility, capitalization, book-to-market, or
momentum.  This finding is an empirical contribution to the asset
pricing because employee's remuneration has not been accounted for in
so far, while it is a determinant element in social sciences including
labor economics, sociology or management.  These various strands of
literature show that strong attention should be paid to wages and more
generally to labor decisions that are likely to affect firms' value.
The economic interpretation of our key finding is mainly based on a
rational explanation of the remuneration anomaly: wages and employee
performance are positively correlated.  This argument is overall
supported by the efficiency wage theory, which claims that rising
wages is the best way to increase output per employee because it links
pecuniary incentives to employee performance.  But it is also
supported by several studies highlighting the prominent role of
operating leverage as a main source of riskiness of equity returns
that is comparable in magnitude to financial leverage.

For this purpose, we introduce an original methodology, coined
``Factor Correlation Level'' (FCL), to build indicator-based factors.
The FCL describes the ability of stocks within the factor to move in a
common way and thus reflects the common risk level underpinning each
factor.  The FCL methodology is a theoretical contribution to the
asset pricing literature.  Indeed, it allows ordering the factors
according to their capacity of taking into account the variability of
stocks.  This ranking can help fund managers to select the most
important factors to set up an asset pricing model and well balanced
portfolios.  The FCL approach is an alternative to the common practice
in asset pricing studies where factor selection depends on several
statistical criteria that do not necessarily convey the same
information.

Implications of this work are important, numerous and go far beyond
asset pricing literature.  A first investment style implication of our
finding is that the companies that pay better should overperform their
competitors by $2.42\%$ per year.  In other words, a market neutral
investment style arbitrage strategy based on the remuneration anomaly
would likely deliver positive returns.  A second economics implication
is that a company might operate better if it could attract the best
human resources while maintaining the company as competitive as
possible by keeping only those employees who are productive.  While we
find that a company that pays too much its shareholders, pays less to
its employees according to the negative correlation between
remuneration and dividend factors, attention should be brought by top
managers to this trade-off between equity capital and labor
remuneration.  A third research implication is that our new
methodology suggests the following ranking for the European stocks
according to their respective FCLs: low-volatility (1.73),
capitalization (1.72), momentum (1.41), sales-to-market (1.22),
liquidity (1.19), book-to-market (1.13), dividend (1.09), leverage
(1.07), remuneration (0.99), and cash (0.92).  In particular, the
low-volatility factor, which is excluded from the FF approach, is the
next most important component following the market factor (i.e., the
stock index).  The remuneration factor is comparable to the
book-to-market factor and thus not negligible.  We conclude that a
five factor model should encapsulate the first five anomalies ordered
by their FCL.

\appendix
\section{Supersectors}
\label{sec:supersectors}

Following the Global Industry Classification Standard (GICS), we
constructed six supersectors as summarized in Table
\ref{tab:supersectors}.  This redistribution has been performed
manually and has aimed at minimizing intrasector correlations and at
obtaining an almost equal number of stocks in each supersector.  We
emphasize that final portfolios include the stocks from all
supersectors, i.e., this redistribution is only an intermediate
technical step to improve the factors.

\begin{table}
\begin{center}
\begin{tabular}{| l | l |}  \hline
1 & Food \& Staples Retailing   \\
  & Food, Beverage \& Tobacco   \\
  & Health Care Equipment \& Services \\
  & Household \& Personal Products \\
  & Pharmaceuticals, Biotechnology \& Life Sciences \\  \hline
2 & Banks \\
  & Diversified Financials \\
  & Insurance \\  \hline
3 & Consumer Durables \& Apparel  \\
  & Consumer Services \\  
  & Media \\
  & Retailing \\  \hline
4 & Materials  \\
  & Real Estate \\   \hline
5 & Energy \\
  & Transportation \\
  & Utilities \\  \hline
6 & Automobiles \& Components  \\
  & Capital Goods  \\
  & Commercial \& Professional Services \\
  & Software \& Services \\
  & Technology Hardware \& Equipment \\
  & Telecommunication Services \\  \hline
\end{tabular}
\end{center}
\caption{
~Six supersectors that we used to split stocks and to construct the
indicator-based factors (from the FACTSET database).  Note that we
mixed very different industries to have 6 supersectors with
approximately the same number of stocks.  Even if different industries
were grouped randomly into six supersectors, we show in Appendix
\ref{sec:FF} that our methodolody would reduce significantly the
sectorial risk of different factors.}
\label{tab:supersectors}
\end{table}

\section{Comparison with FF approach}
\label{sec:FF}

In order to highlight the advantages of our methodology as compared to
the standard FF approach, it is instructive to consider {\it
incremental} transformations from one method to the other.  In this
way, one can analyze the respective roles of several proposed
improvements.  For this purpose, we implement the standard FF approach
and its progressive modifications. 

\begin{itemize}
\item 
A0 (the standard FF approach): According to Table I from
\citet{Fama15b}, stocks are subdivided two groups of small (below
median) and large (above median) capitalization.  Within each of two
groups, assets are ordered according to the chosen indicator (e.g.,
remuneration) and then split into three subgroups (top, medium and
bottom $33\%$).  The related portfolio is constructed by buying the
top $33\%$ and selling the bottom $33\%$ assets from the sorted list
with equal weights.  Such prepared two portfolios (for small and large
capitalization groups) are then merged into a single FF portfolio.  To
be comparable with our methodology, the portfolio is rebalanced on
daily basis (note that the original FF approach stipulated monthly
rebalancing).  The constucted portfolio is delta-neutral.

\item
A1: The same rules as A0 except for buying top $15\%$ and selling
bottom $15\%$ assets (as in our methodology);

\item 
A2: The same rules as A1 except that the splitting into small
and large capitalization groups is withdrown;

\item
A3: The same rules as A2 except that we add sectorial and geographical
constraints as in our methodology.  In other words, assets are split
into 6 supersectors (see Appendix \ref{sec:supersectors}), the
portfolio construction is performed individually for each supersector
and then the obtained portfolios are merged.  In addition, we
normalize the chosen indicator (e.g., remuneration) by the median per
country to correct for geographical biases;

\item
A4: The same rules as A3 except that equal weights are replaced by
volatility-based weights as in our methodology;

\item
A5: The same rules as A4 except that the volatility-based weights are
rescaled by factors $\mu_{\pm}$ to get beta-neutral portfolios (beta's
are estimated throuh a standard methodology);

\item
A6 (our methodology): The same rules as A5 except that a standard
volatility and beta estimations (by exponential moving averages) are
replaced by the reactive volatility model.
\end{itemize}

					
Each of these seven approaches (A0, ..., A6) has been applied to both
U.K. and European universes.  We computed the mean return and
volatility of ten factor-based portfolios introduced in this paper.
To be closer to the standard Fama and French framework, we present
results on {\it monthly} basis, in contrast to the main text, in which
daily basis was used.  Table \ref{tab:FF} recapitulates the main
findings for the European universe (similar results were obtained for
the U.K. universe, available upon request).

\begin{table}
\begin{center}
\footnotesize
\begin{tabular}{|l|l|l|l|l|l|l|l|l|l|l|l|}  \hline
	&	& Div.	& Cap.	& Low	& Mom.	& Liq.	& Lev.   & Sales.& Book. & Rem.	& Cash.  \\  \hline
\multirow{3}{3mm}[-1mm]{\begin{turn}{90}A0\end{turn}}
	&Mean	&0.35\%	&-0.93\%& 0.30\%&-0.64\%& 0.68\%& 0.02\% &-0.46\%&-0.33\%&-0.03\%& 0.35\% \\
	&Std    &3.20\%	& 0.56\%& 4.90\%& 5.72\%& 3.92\%& 2.87\% & 3.16\%& 3.16\%& 2.04\%& 1.99\% \\
	&t-stat	&1.46   &-22.40 & 0.83  &-1.49  & 2.32  & 0.10   &-1.97  &-1.40  &-0.18  & 2.37   \\  \hline
\multirow{3}{3mm}[-1mm]{\begin{turn}{90}A1\end{turn}}
	&Mean	&0.37\% &-0.92\%& 0.34\%&-0.79\%& 0.75\%& 0.07\% &-0.53\%&-0.42\%&-0.02\%& 0.32\% \\
	&Std    &3.15\% & 0.44\%& 4.81\%& 5.52\%& 3.77\%& 2.80\% & 2.98\%& 3.04\%& 1.98\%& 1.97\% \\
	&t-stat	&1.58   &-27.98 & 0.95  &-1.91  & 2.66  & 0.32   &-2.40  &-1.87  &-0.12  & 2.17   \\  \hline
\multirow{3}{3mm}[-1mm]{\begin{turn}{90}A2\end{turn}}
	&Mean	&0.37\% &-1.12\%&-0.21\%&-0.49\%& 0.31\%&-0.23\% &-0.49\%&-0.27\%&-0.07\%& 0.38\% \\
	&Std	&3.41\% & 1.39\%& 4.80\%& 6.01\%& 3.85\%& 2.54\% & 3.18\%& 3.47\%& 2.00\%& 1.96\% \\
	&t-stat	&1.45   &-10.82 & -0.59 & -1.09 & 1.07  & -1.20  & -2.09 & -1.05 &-0.44  & 2.62   \\  \hline
\multirow{3}{3mm}[-1mm]{\begin{turn}{90}A3\end{turn}}
	&Mean	&0.41\% &-0.96\%&-0.19\%&-0.61\%& 0.31\%&-0.21\% &-0.40\%&-0.39\%& 0.00\%& 0.39\% \\
	&Std    &2.65\% & 1.17\%& 3.85\%& 4.99\%& 3.35\%& 2.31\% & 3.05\%& 2.60\%& 1.91\%& 1.69\% \\
	&t-stat	&2.06   &-10.97 & -0.68 & -1.63 & 1.22  & -1.23  & -1.77 &-2.03  & 0.02  & 3.11   \\  \hline
\multirow{3}{3mm}[-1mm]{\begin{turn}{90}A4\end{turn}}
	&Mean	&0.41\% &-0.96\%&-0.19\%&-0.60\%& 0.30\%&-0.21\% &-0.41\%&-0.40\%& 0.00\%& 0.40\% \\
	&Std    &2.65\% & 1.17\%& 3.85\%& 4.98\%& 3.34\%& 2.31\% & 3.05\%& 2.59\%& 1.91\%& 1.68\% \\
	&t-stat & 2.06  &-10.97 & -0.68 & -1.62 & 1.22  &-1.19   & -1.79 & -2.06 & 0.03  & 3.17   \\  \hline
\multirow{3}{3mm}[-1mm]{\begin{turn}{90}A5\end{turn}}
	&Mean	& 0.41\%&-1.16\%&-0.86\%&-0.11\%&-0.34\%&-0.46\% & 0.02\%&-0.08\%& 0.22\%& 0.25\% \\
	&Std    & 2.09\%& 1.97\%& 1.90\%& 3.34\%& 2.37\%& 1.58\% & 1.95\%& 1.94\%& 1.53\%& 1.61\% \\
	&t-stat & 2.61  & -7.88 & -6.04 & -0.43 & -1.94 & -3.94  & 0.13  &-0.58  & 1.92  & 2.06   \\  \hline
\multirow{3}{3mm}[-1mm]{\begin{turn}{90}A6\end{turn}}
	&Mean	& 0.45\%&-1.17\%&-0.82\%&-0.16\%&-0.36\%&-0.40\% &-0.03\%&-0.10\%& 0.19\%& 0.24\% \\
	&Std    & 2.05\%& 1.91\%& 1.94\%& 3.33\%& 2.44\%& 1.59\% & 2.06\%& 2.00\%& 1.50\%& 1.60\% \\
	&t-stat	& 2.94  & -8.19 & -5.63 & -0.66 & -2.00 & -3.36  & -0.22 &-0.66  & 1.73  & 1.98   \\  \hline
\end{tabular}
\end{center}
\caption{
~Progressive evaluation of factor performances with incremental
transition from the FF approach (A0, top) to our methodology (A6,
bottom).  For each factor, we present mean monthly return (Mean) and
volatility (Std), as well as their ratio (t-stat).}
\label{tab:FF}
\end{table}

As expected, the change of quantiles (passage from the standard A0
approach to A1) almost does not affect the results.  Similarly, a
standard volatility/beta estimator and the reactive volatility/beta
model lead to similar results (passage from A5 to A6).  The most
significant changes are observed when passing from A2 to A3 and from
A4 to A5.

$\bullet$ In the former case, adding the sectorial constraints (see
Appendix \ref{sec:supersectors}) reduces sectorial biases and allows
one to better capture the indicator-based factors.  To illustrate this
point, let us suppose that remuneration is very high in the energy
industry and is low (at approximately the same level) in all other
industries.  If there was no sectorial constraint, the remuneration
factor would be long on the energy industry and short in all other
industries.  In other words, it would be $100\%$ invested in energy,
with eventual high risks.  In turn, the sectorial constraint reduces
this risk by approximately 1/6 because the strong concentration on
energy only remains in the 5th supersector while investments in other
industries are necessarily imposed for other supersectors.  For
instance, if the annualized sectorial volatility is $12\%$, such an
enforced diversification would reduce it to $2\%$ on an annualized
basis.

$\bullet$ In the latter case, we switch from the delta-neutral to
beta-neutral portfolios, i.e., we (partly) remove correlations with
the stock market index.  We evoque two possible origins to rationalize
the significant decrease of volatility when passing from A4 to A5.
First, if we suppose that stock beta's follow a distribution with
standard deviation $s_\beta$, the average aggregated beta of a random
delta-neutral factor built with $2 \times 15\% \times 500 = 150$
stocks would be $0$, while its standard deviation would be $2s_\beta
/\sqrt{150} \approx 16\% s_\beta \approx 6\%$, where we estimated
$s_\beta \approx 0.37$ from our data.  As a consequence, the
volatility added by the random exposure to the market index is around
$6\% \times \sigma_m \approx 1.2\%$ on an annualized basis, where
$\sigma_m \approx 21\%$ is the annualized volatility of the market
index.  Second, our construction of beta-neutral portfolio reduces
their leverage to ensure Eq. (\ref{eq:beta_neutral}).  Consequently,
smaller investments lead to smaller volatility, as compared to the
Fama and French construction with a constant investment.

One also observes that volatilities of factors progressively diminish
when passing from A0 to A6.  This observation indicates that our
modifications better withdraw other common risks and manage to
concentrate on the risk of interest.

Looking more specifically to the remuneration factor, one can observe
a significant increase of t-stat, from $-0.18$ (insignificant) to
$1.73$ (significant), when passing from the standard FF approach (A0)
to our methodology (A6).  In other words, {\bf implementing the above
improvements allowed us to level up the remuneration factor from noise
to a small but significant anomaly.}

We complete this Appendix by the following general remark.  The
variability of results presented in Table \ref{tab:FF} indicates their
dependence on a chosen data analysis method and its parameters.  The
methodology plays therefore the crucial role, especially when dealing
with small anomalies such as remuneration.  This highlights the
advantage of our method that enabled to detect and quantify such small
features in the market behavior.  At the same time, our methodology
remains robust against some changes in construction of factors, such
as replacing conventional volatility estimator by reactive volatility
model, using volatility renormalized weights, or changing daily to
monthly returns.




\begin{thebibliography}{99}


\bibitem[Abowd\etal(1999)]{Abowd99}      
							Abowd, J.M., Kramarz, F., and Margolis,	D.M.  1999.
							High wage workers and high wage	firms.  
							{\it Econometrica} 67, 251-333.


\bibitem[Adams(1963)]{Adams63}      
							Adams, J.S. 1963. 
							Toward an understanding of inequity. 
							{\it Journal of Abnormal and Social Psychology} 67, 422-436.

\bibitem[Akerlof(1982)]{Akerlof82}      
							Akerlof, G.A. 1982. 
							Labour contracts as a partial gift exchange. 
							{\it Quarterly Journal of Economics} 97, 543-69.


\bibitem[Akerlof and Yellen(1990)]{Akerlof90}      
							Akerlof, G.A. and Yellen, J.L. 1990. 
							The fair wage-effort hypothesis and unemployment. 
							{\it Quarterly Journal of Economics} 105, 255-283.

\bibitem[Alchian and Demsetz(1972)]{Alchian72}      
							Alchian, A. and Demsetz, H. 1972. 
							Production, information costs, and economic organization. 
							{\it American Economic Review} 62, 777-795.

\bibitem[Allez and Bouchaud(2012)]{Allez12}       
				                        Allez R. and Bouchaud J.-P., 2012.
				                        Eigenvector dynamics: General theory and some applications.
				                        {\it Physical Reviews E} 86, 046202.


\bibitem[Andersen\etal(2000)]{Andersen00}    
				                        Andersen T. G., Bollerslev T., Diebold F. X., and Labys P., 2000.
				                        Exchange rate returns standardized by realized volatility are (nearly) gaussian.
				                        {\it Multinational Finance Journal} 4, 159-179.

\bibitem[Ang\etal(2006)]{Ang06}    
				                        Ang, A., Hodrick,  R., Xing, Y., and Zhang., X., 2006.
				                        The cross-section of volatility and expected returns.
				                        {\it Journal of Finance} 61, 259-299.
														

\bibitem[Avramov and Chordia(2006)]{Avramov06}      
				                        Avramov, D., Chordia, T. 2006. 
				                        Asset pricing models and financial market anomalies. 
				                        {\it Review of Financial Studies} 19, 1001-1040.

\bibitem[Babecky\etal(2012)]{Babecky12}
							Babecky, J., Du Caju, P., Kosma, T., Lawless, M., Messina, J., and Room, T. 2012.
							How do European firms adjust their labour costs when nominal wages are rigid? 
							{\it Labour Economics} 19, 732-801.

\bibitem[Baker(1992)]{Baker92}      
							Baker, G. 1992. 
							Incentive contracts and performance measurement. 
							{\it Journal of Political Economy} 100, 598-614.

\bibitem[Bali\etal(2013)]{Bali13}        
				                        Bali, T.G., Cakici, N., and Fabozzi, F.J. 2013. 
				                        Book-to-market and the cross-section of expected returns in international stock markets. 
				                        {\it Journal of Portfolio Management} 39, 101-115.

\bibitem[Banz(1981)]{Banz81}        
				                        Banz, R.W. 1981. 
				                        The relationship between return and market value of common stocks. 
				                        {\it Journal of Financial Economics} 9, 3-18.

\bibitem[Bebchuk\etal(2002)]{Bebchuk02}     
				                        Bebchuk, L., Fried, J.M., and Walker, D.I. 2002. 
				                        Managerial power and rent extraction in the design of executive compensation. 
				                        {\it University of Chicago Law Review} 69, 751-846.

\bibitem[Belfield\etal(2004)]{Belfieldk04}     
							Belfield, C.R., and Wei, X. 2004. 
							Employer size-wage effects : evidence from matched employer-employee survey data in the UK. 
							{\it Applied Economics} 36, 185-193.

\bibitem[Belo\etal(2014)]{Belo14}        
							Belo, F., Lin, X., and Bazdresch, S. 2014.	
							Labor hiring, investment and stock return predictability in the cross section. 
							{\it Journal of Political Economy} 122, 129-177.

\bibitem[Black\etal(1972)]{Black72}      
							Black, F., Jensen, M.C., and Scholes, M. 1972. 
							The capital asset pricing model: Some empirical tests. 
							{\it Studies in the Theory of Capital Markets}, New York: Praeger.

\bibitem[Blau(1955)]{Blau55}      
							Blau, P.M. 1955. 
							The Dynamics of bureaucracy: A study of interpersonal relations in two government agencies.
							(Chicago: Chicago University Press).

\bibitem[Bouchaud\etal(2001)]{Bouchaud01b}   
				                        Bouchaud J.-P., Matacz A., and Potters M., 2001.
				                        Leverage effect in financial markets: The retarded volatility model.
				                        {\it Physical Review Letters}  87, 1-4.

\bibitem[Campbell and Vuolteenaho(2004)]{Campbell04}      
							Campbell, J.Y., and Vuolteenaho, T. 2004.
							Bad beta, good beta.
							{\it American Economic Review} 94,  1249-1275.

\bibitem[Carhart(1997)]{Carhart97}
				                        Carhart, M. 1997. 
				                        On persistence in mutual fund performance. 
				                        {\it Journal of Finance} 52, 57-82.

\bibitem[Carlson\etal(2004)]{Carlson04}        
							Carlson, M., Fisher, A., and Giammarino, R. 2004.	 
							Corporate investment and asset price dynamics: Implications for the cross-section of returns. 
							{\it Journal of Finance} 59, 2577-2603.

\bibitem[Cederburg\etal(2015)]{Cederburg15}   
				                        Cederburg, S., Davies, P., O'Doherty, M. 2015. 
				                        Asset-pricing anomalies at the firm level. 
				                        {\it Journal of Econometrics} 186, 113-128.

\bibitem[Chen\etal(1986)]{Chen86}        
				                        Chen, N.F., Roll, R., and Ross, S.A. 1986. 
				                        Economic forces and the stock market. 
				                        {\it Journal of Business} 59, 383-403.

\bibitem[Cheng\etal(2015)]{Cheng15}       
				                        Cheng, I.-H., Hong, H. and Scheinkman, J.E.A. 2015. 
				                        Yesterday's heroes: Compensation and risk at financial firms. 
				                        {\it Journal of Finance} 70, 839-879.


\bibitem[Cochrane(2005)]{Cochrane05}      
							Cochrane, J. 2005. 
							Asset pricing. Revised edition. 
							{\it Princeton University Press}.

\bibitem[Danthine and Donaldson(2002)]{Danthine02}        
							Danthine, J., and Donaldson, J.	2002. 
							Labor relations and asset pricing. 
							{\it Review of Economic Studies} 69, 41-64.

\bibitem[Deci\etal(1999)]{Deci99}      
							Deci, E.L., Koestner, R., and Ryan, R. 1999. 
							A meta-analytic review of experiments examining the effects of extrinsic rewards on intrinsic motivation. 
							{\it Psychological Bulletin} 125, 627-668.

\bibitem[Donangelo(2014)]{Donangelo14}        
							Donangelo, A. 2014. 
							Labor mobility: Implications for asset pricing. 
							{\it Journal of Finance} 3, 1321-1346.

\bibitem[Fama and MacBeth(1973)]{Fama73}      
							Fama, E.F., MacBeth, J.D. 1973. 
							Risk, return, and equilibrium: Empirical tests. 
							{\it Journal of Political Economy} 81, 607-636.

\bibitem[Fama(1980)]{Fama80}
							Fama, E. 1980. 
							Agency problems and the theory of the firm. 
							{\it Journal of Political Economy} 88, 288-307.

\bibitem[Fama and French(1992)]{Fama92}
							Fama, E., and French, K.R. 1992. 
							The cross-section of expected stock returns. 
							{\it Journal of Finance} 47, 427-465.

\bibitem[Fama and French(1993)]{Fama93}        
				                        Fama, E.F., and French, K.R. 1993. 
				                        Common risk factors in the returns on stocks and bonds. 
				                        {\it Journal of Financial Economics} 33, 3-56.

\bibitem[Fama and French(1996)]{Fama96}        
				                        Fama, E.F., and French, K.R. 1996. 
				                        Multifactor explanations of asset pricing anomalies. 
				                        {\it Journal of Finance} 51, 55-84.

\bibitem[Fama and French(1998)]{Fama98}        
				                        Fama, E.F., and French, K.R. 1998. 
				                        Value vs. Growth: The international evidence. 
				                        {\it Journal of Finance} 53, 1975-1999.

\bibitem[Fama and French(2008)]{Fama08}        
				                        Fama, E.F., and French, K.R. 2008. 
				                        Dissecting anomalies. 
				                        {\it Journal of Finance} 63, 1653-1678.

\bibitem[Fama and French(2012)]{Fama12}        
				                        Fama, E.F., and French, K.R. 2012. 
				                        Size, value, and momentum in international stock returns. 
				                        {\it Journal of Financial Economics} 105, 457-472.

\bibitem[Fama and French(2015)]{Fama15}        
				                        Fama, E.F., and French, K.R. 2015. 
				                        A five factor model. 
				                        {\it Journal of Financial Economics} 116, 1-22.

\bibitem[Fama and French(2015)]{Fama15b}        
				                        Fama, E.F., and French, K.R. 2015. 
							Dissecting Anomalies with a Five-Factor Model.
							Working paper.



\bibitem[Favilukis and Xiaoji(2016)]{Favilukis16}        
							Favilukis, J., and Xiaoji L.  2016. 
							Wage rigidity:  A quantitative solution to several asset pricing puzzles. 
							{\it Review of Financial Studies} 29, 148-192.

\bibitem[Fehr and Falk(2002)]{Fehr02}      
							Fehr, E. and Falk, A. 2002. 
							Psychological foundations of incentives. 
							{\it European Economic Review} 46, 687-724.

\bibitem[Ferguson and Shockley(2003)]{Ferguson03}    
				                        Ferguson, M.F., and Shockley, R.L. 2003. 
				                        Equilibrium anomalies. 
				                        {\it Journal of Finance} 58, 2549-2580.

\bibitem[Fu(2009)]{Fu09}
							Fu, F., 2009. 
							Idiosyncratic risk and the cross-section of expected stock returns.
							{\it Journal of Financial Economics} 91, 24-37.

\bibitem[Gabaix and Landier(2008)]{Gabaix08}      
				                        Gabaix, X., and Landier, A. 2008. 
				                        Why has CEO pay increased so much? 
				                        {\it Quarterly Journal of Economics} 123, 49-100.

\bibitem[Gandhi and Lustig(2015)]{Gandhi15}      
				                        Gandhi, P. and Lustig, H. 2015. 
				                        Size anomalies in U.S. bank stock returns. 
				                        {\it Journal of Finance} 70, 733-768.

\bibitem[Gaver and Gaver(1995)]{Gaver95}       
				                        Gaver, J., and  Gaver, K.M. 1995. 
				                        Compensation policy and the investment opportunity set. 
				                        {\it Financial Management} 24, 19-32.

\bibitem[Gibbons\etal(1989)]{Gibbons89}                 Gibbons, M.R., Ross, S.A., and Shanken, J. 1989.
                                                        A test of the efficiency of a given portfolio.
							{\it Econometrica} 57, 1121-1152. 

\bibitem[Gibbons and Murphy(1992)]{Gibbons92}
							Gibbons, M.R., and Murphy, K.J. 1992. 
							Optimal incentive contracts in the presence of career concerns: Theory and evidence.
							{\it Journal of Political Economy}  100, 468-505.

\bibitem[Gneezy\etal(2011)]{Gneezy11}      
				                        Gneezy, U., Meier, S. and Rey-Biel, P. 2011. 
				                        When and why incentives (don't) work to modify behavior. 
				                        {\it Journal of Economic Perspectives} 25, 191-209.

\bibitem[Gomez\etal(2015)]{Gomez15}      
							Gomez, J-P., Priestley, R., and Zapatero, F. 2015.
							Labor income, relative wealth concerns, and the cross-section of stock returns. 
							Forthcoming, {\it Journal of Financial and Quantitative Analysis}.

\bibitem[Gourio(2007)]{Gourio07}        
							Gourio, F. 2007. 
							Labor leverage, firms heterogeneous sensitivities to the business cycle, and the cross-section of returns. 
							Working paper, Boston University

\bibitem[Graham\etal(2012)]{Graham12}      
				                        Graham, J.R., Li, S. and Qiu, J. 2012. 
				                        Managerial attributes and executive compensation. 
				                        {\it Review of Financial Studies} 25, 144-186.

\bibitem[Green and Heywood(2008)]{Green08}       
				                        Green, C., and Heywood, J.S. 2008. 
				                        Does performance pay increase job satisfaction? 
				                        {\it Economica} 75, 710-728.

\bibitem[Grinblatt and Moskowitz(2004)]{Grinblatt04}   
				                        Grinblatt, M., and Moskowitz, T. 2004. 
				                        Predicting stock price movements from past returns: The role of consistency and tax-loss selling. 
				                        {\it Journal of Financial Economics} 71, 541-579.

\bibitem[Hall and Murphy(2003)]{Hall03}        
				                        Hall, B.J., and Murphy, K.J. 2003. 
				                        The trouble with stock options. 
				                        {\it Journal of Economic Perspectives} 17, 49-70.

\bibitem[Harvey\etal(2015)]{Harvey15a}     
							Harvey, C.R., Liu, Y., and Zhu, H. 2015.
							... and the cross-section of expected returns.
							Forthcoming, {\it Review of Financial Studies}.

\bibitem[Harvey and Liu(2015)]{Harvey15b}     
							Harvey, C.R., Liu, Y. 2015.
							Lucky factors.
							Working paper, University of Duke.

\bibitem[Holmstrom(1999)]{Holmstrom99}
							Holmstrom, B. 1999. 
							Managerial incentive problems: A dynamic perspective. 
							{\it Review of Economic Studies} 66, 169-182.

\bibitem[Hou\etal(2015)]{Hou15}
							Hou, K., Xue, C. and Zhang, L. 2014.
							Digesting anomalies: An investment approach.
							{\it Review of Financial Studies} 28, 650-705.

\bibitem[Jegadeesh and Titman(1993)]{Jegadeesh93}   
				                        Jegadeesh, N., and Titman, S. 1993. 
				                        Returns to buying winners and selling losers: Implications for stock market efficiency. 
				                        {\it Journal of Finance} 48, 65-91.

\bibitem[Jensen\etal(2004)]{Jensen04}      
				                        Jensen, M., Murphy, K.J. and Wruck, E. 2004. 
				                        Remuneration: Where we've been, how we got to here, what are the problems, and how to fix them. 
				                        Mimeo, Harvard University.

\bibitem[Jordan and Riley(2013)]{Jordan13}      
				                        Jordan, B.D. and Riley, T.B. 2013. 
				                        Dissecting the low volatility anomaly. 
				                        Working paper, University of Kentucky.

\bibitem[Kuehn\etal(2013)]{Kuehn13}        
							Kuehn, L-A., Petrosky-Nadeau, N., and Zhang, L. 2013.	 
							An equilibrium asset pricing model with labor market search. 
							Working paper, NBER.

\bibitem[Lallemand\etal(2007)]{Lallemand07}      
							Thierry Lallemand, T., Plasman, R., and Rycx, F. 2007. 
							The establishment-size wage premium: Evidence from European countries.
							{\it Empirica} 34, 427-451.


\bibitem[Laloux\etal(1999)]{Laloux99}      
				                        Laloux L., Cizeau P., Bouchaud J.-P., and Potters M. 1999.
				                        Noise dressing of financial correlation matrices.
				                        {\it Physical Review Letters} 83, 1467.

\bibitem[Lazear(2000)]{Lazear00}      
				                        Lazear, E.P. 2000. 
				                        Performance pay and productivity. 
				                        {\it American Economic Review} 90, 1346-1361.

\bibitem[Liu and Zhang(2008)]{Liu08}         
				                        Liu, X., and L. Zhang, 2008, 
				                        Momentum profits, factor pricing, and macroeconomic risk. 
				                        {\it Review of Financial Studies} 21, 2417-2448.

\bibitem[McLean and Pontiff(2015)]{Mclean15}
							McLean, R.D. and Pontiff, J. 2015, 
							Does academic research destroy stock return predictability? 
							Forthcoming, {\it Journal of Finance}.
											
\bibitem[Monika and Yashiv(2007)]{Monika07}        
							Monika, M., and Yashiv, E. 2007. 
							Labor and the market value of the firm. 
							{\it American Economic Review} 97, 1419-1431.

\bibitem[O'Byrne and Young(2010)]{OByrne10}      
				                        O'Byrne, S.F. and Young, S.D. 2010. 
				                        What investors need to know about executive pay. 
				                        {\it Journal of Investing} 19, 36-44.

\bibitem[Ochoa(2013)]{Ochoa13}        
							Ochoa, M. 2013. 
							Volatility, labor heterogeneity and asset prices. 
							Working paper, Federal Reserve Board of Washington D.C.

\bibitem[Oi and Idson(1999)]{Oi99}      
							Oi, W.Y. and Idson, T.L. 1999.
							Firm size and wages.
							{\it Handbook of labor economics} 3, 2165-2214.

\bibitem[Idson and Oi(1999)]{Idson99}      
							Idson, T.L., and Oi, W.Y. 1999. 
							Workers are more productive in large firms. 
							{\it American Economic Review} 89, 104-108.

\bibitem[Plerou\etal(1999)]{Plerou99}      
				                        Plerou V., Gopikrishnan P., Rosenow B., Amaral L. A. N., and Stanley H.E., 1999.
				                        Universal and nonuniversal properties of cross correlations in financial time series.
				                        {\it Physical Review Letters} 83, 1471.

\bibitem[Plerou\etal(2002)]{Plerou02}      
				                        Plerou V., Gopikrishnan P., Rosenow B., Amaral L. A. N., Guhr T., and Stanley H. E., 2002.
				                        Random matrix approach to cross correlations in financial data.
				                        {\it Physical Reviews E} 65, 066126.

\bibitem[Potters\etal(2005)]{Potters05}     
				                        Potters M., Bouchaud J.-P., and Laloux L., 2005.
				                        Financial applications of random matrix theory: Old laces and new pieces.
				                        {\it Acta Physica Polonica B} 36, 2767-2784.

\bibitem[Rynes\etal(1983)]{Rynes83}      
							Rynes, S., Schwab, D., and Heneman, H. 1983. 
							The role of pay and market pay variability in job application decisions. 
							{\it Organizational Behavior and Human Performance} 31, 353-364.

\bibitem[Rynes\etal(2004)]{Rynes04}      
							Rynes, S., Gerhart, B., and Minette, K. 2004. 
							The importance of pay in employee motivation: Discrepancies between what people say and what they do. 
							{\it Human Resource Management} 43, 381-394.

\bibitem[Santos and Veronesi(2006)]{Santos06}        
							Santos, T., and Veronesi, P. 2006. 
							Labor income and predictable stock returns. 
							{\it Review of Financial Studies} 19, 1-44.

\bibitem[Shapiro and Stiglitz(1984)]{Shapiro84}      
							Shapiro, C., and Stiglitz, J.E. 1984. 
							Equilibrium unemployment as a worker discipline device.
							{\it American Economic Review} 74, 433-444.

\bibitem[Stiglitz(1981)]{Stiglitz81}      
							Stiglitz, J.E. 1981. 
							Alternative theories of wage determination and unemployment: The efficiency wage model. 
							Working paper, University of Princeton.


\bibitem[Schwert(2003)]{Schwert03}     
				                        Schwert, G.W. 2003. 
				                        Anomalies and market efficiency. 
				                        Chapter 15 in Handbook of the economics of finance. 
							eds. G. Constantinides, M. Harris, and R. M. Stulz, North-Holland, 937-972.

\bibitem[Valeyre\etal(2013)]{Valeyre13}     
				                        Valeyre S., Grebenkov D. S., Aboura S., and Liu Q., 2013.
				                        The reactive volatility model.
				                        {\it Quantitative Finance} 13, 1697-1706.

\bibitem[Wang\etal(2011)]{Wang11}        
				                        Wang D., Podobnik B., Horvatic D., and Stanley H. E., 2011.
				                        Quantifying and modeling long-range cross correlations in multiple time series
							with applications to world stock indices.
				                        {\it Physical Reviews E} 83, 046121.

\bibitem[Winter-Ebmer\etal(1999)]{Winter99}
							Winter-Ebmer, R., and Zweimuller, J. 1999. 
							Firm-size wage differentials in Switzerland: Evidence from job-changers.
							{\it American Economic Review} 89, 89-93.

\bibitem[Yellen(1984)]{Yellen84}      
							Yellen, J.L. 1984. 
							Efficiency wage models of unemployment. 
							{\it American Economic Review} 74, 200-205.



												
\end{thebibliography}
\end{document}